\documentclass[final,authoryear,3p,times,twocolumn]{elsarticle}
\usepackage{nicefrac}
\usepackage{mathdots}
\usepackage{bm}
\usepackage{mathtools}
\usepackage{amsmath,amssymb}
\usepackage{amsfonts}
\usepackage{algorithm}
\usepackage{algorithmic}
\usepackage{algorithmwh}
\usepackage{float}
\usepackage{xspace}
\usepackage{verbatim}
\usepackage[T1]{fontenc}
\usepackage{prettyref}
\usepackage{url}
\usepackage{graphicx}
\usepackage{xcolor}
\usepackage{array}
\usepackage{pifont}
\usepackage{caption}
\usepackage{subcaption}
\usepackage{capt-of}
\usepackage{enumitem}
\usepackage{rotating}
\usepackage{color}
\usepackage{array}
\usepackage{longtable}
\usepackage{calc}
\usepackage{multirow}
\usepackage{hhline}
\usepackage{ifthen}
\usepackage{lscape}
\usepackage{lettrine}
\usepackage[active]{srcltx}
\usepackage{cite}
\usepackage{hyperref}

\usepackage{endnotes}
\let\footnote=\endnote

\usepackage[T1]{fontenc}
\usepackage{mwe}
\usepackage{caption}
\usepackage{subcaption}
\usepackage{newfloat}
\DeclareFloatingEnvironment[%
 fileext=loe,
 name=Exhibit
]{exhibit}

\journal{arXiv}
\begin{document}
\begin{frontmatter}
\title{Hidden Markov Models Applied To Intraday Momentum Trading With Side Information}
\author[label1]{Hugh Christensen\corref{cor1}}
\ead{hlc54@cam.ac.uk}
\cortext[cor1]{Corresponding author.}
\address[label1]{Signal Processing and Communications Laboratory, Engineering Department, Cambridge University, CB2 1PZ, UK}

\author[label2]{Richard Turner}
\ead{ret26@cam.ac.uk}
\address[label2]{Machine Learning Group, Engineering Department, Cambridge University, CB2 1PZ, UK}

\author[label1]{Simon Godsill}
\ead{sjg30@cam.ac.uk}

\begin{abstract}
A Hidden Markov Model for intraday momentum trading is presented which specifies a latent momentum state responsible for generating the observed securities' noisy returns. Existing momentum trading models suffer from time-lagging caused by the delayed frequency response of digital filters. Time-lagging results in a momentum signal of the wrong sign, when the market changes trend direction. A key feature of this state space formulation, is no such lagging occurs, allowing for accurate shifts in signal sign at market change points.  The number of latent states in the model is estimated using three techniques, cross validation, penalized likelihood criteria and simulation based model selection for the marginal likelihood. All three techniques suggest either 2 or 3 hidden states. Model parameters are then found using Baum-Welch and Markov Chain Monte Carlo, whilst assuming a single (discretized) univariate Gaussian distribution for the emission matrix. Often a momentum trader will want to condition their trading signals on additional information. To reflect this, learning is also carried out in the presence of side information. Two sets of side information are considered, namely a ratio of realized volatilities and intraday seasonality. It is shown that splines can be used to capture statistically significant relationships from this information, allowing returns to be predicted.   An Input Output Hidden Markov Model is used to incorporate these univariate predictive signals into the transition matrix, presenting a possible solution for dealing with the signal combination problem.  Bayesian inference is then carried out to predict the securities $t+1$ return using the forward algorithm.  The model is simulated on one year's worth of e-mini S\&P500 futures data at one minute sampling frequency, and it is shown that pre-cost the models have a Sharpe ratio in excess of 2.0. Simple modifications to the current framework allow for a fully non-parametric model with asynchronous prediction.
\end{abstract}

\begin{keyword}
Bayesian inference, trend following, high frequency futures trading, quantitative finance.
\end{keyword}

\end{frontmatter}

\section{Introduction}
\label{mainIntro}
An intraday momentum trading strategy is presented, consisting of a Hidden Markov Model (HMM) framework that has the ability to use side information from external predictors.   The proposed framework is quite general and allows any predictors to be used in conjunction with the momentum model.   An appealing aspect of this model is that all the computationally demanding learning is done \emph{off-line}, allowing for a fast inference phase meaning the model can be applied to high-frequency financial data.

Quantitative trading, namely the application of the scientific method, is now well established in the financial markets. A sub-section of this field is termed algorithmic trading, where algorithms are responsible for the full trade cycle, including the decision of when to buy and sell.   When this process is dependent on the prior behavior of the security, it historically was termed \emph{technical analysis} \citep{Lo2000a}.   Momentum trading (or trend following) falls into this category and is the most popular hedge fund style trading strategy currently used. For example, the largest quantitative hedge funds by assets under management famously trade momentum strategies \citep{EconomistJan2011}.  It can be inferred from this that momentum is the most significant exploitable effect in the financial markets, and as a result of this there is a large body of literature published on the effect \citep{hong1999unified}.  Momentum (or trend) can be defined as the rate of change of price. As a strategy, momentum trading aims to forecast future security returns by exploiting the positive autocorrelation structure of the data.  Once a trend is detected by careful estimation of the mean return (in the presence of noise), it can be predicted.    The most well known trend-following system is that introduced by Gerald Appel in the 1970's, the moving-average convergence-divergence (MACD) \citep{A.Gerald1999}, made famous by the success of a group of traders named the ``turtles'' \citep{Faith2007a}.  The MACD strategy uses the difference between a pair of cascaded low pass filters in parallel to remove noise while estimating the true mean of the rate of change of price \citep{Satchell2002a}.        The reasons for the momentum effect existing are less than clear despite extensive academic research on the subject. Financial data consists of deterministic and stochastic components and both of these components can exhibit trends. Significant trends commonly occur even in data which is generated by a random process, such as geometric Brownian motion \citep{wilmott2006a} and can be explained by the effect of summing random disturbances \citep{lo2001nonB}. Attempting to model such stochastic trends can lead to spurious results.  Deterministic reasons for trends existing are thought to include herding behaviour \citep{Shiller2005a}, supply-and-demand arguments \citep{Johnson2002} and delayed over-reactions that are eventually reversed \citep{jegadeesh1999profitabilityB}.     While there is debate in the academic literature between those that believe the momentum effect is still viable post-transaction costs, for example \citep{jegadeesh1999profitabilityB}, and those that believe the effect has been arbitraged away, for example \citep{Lesmond2004a}, the continued profitability of large momentum trading hedge funds is testament to the enduring nature of the momentum effect.

The motivation for this paper is to apply HMM's to produce a trading algorithm that exploits the momentum effect, and that can be applied to the financial markets in real-time by industry practitioners. The core aim of the paper is to give the algorithm the best predictive performance possible, irrespective of methodology. Application of such work to the financial markets has obvious economic benefits.

The two main innovations presented in this paper are both new and novel applications of existing statistical techniques to an applied problem. No new methodologies are introduced in the paper.  Firstly, the price discovery process of a security is described by a trend term in the presence of noise. This process is fitted into an HMM framework and various means of parameter estimation are inspected. Secondly, momentum traders often want to incorporate other information into their momentum based forecast, the \emph{signal combination problem}, and an IOHMM framework is established to allow this. For both innovations, realistic experiments are conducted (including transaction costs and slippage), results presented and conclusions drawn.

This paper is structured as follows. In Section \ref{hmmIntroSection} HMM's in finance and economics are reviewed and the HMM framework is introduced. In Section \ref{learningSection} the three learning methodologies are presented. In Section \ref{sideInfoSection} two extrinsic predictors are developed and tested, and then in Section \ref{learningWithSideInfo} learning is carried out using this side information. In Section \ref{inferanceSection} our inference algorithm is presented.  In Section \ref{dataSimSection} we present the historical futures data and then simulate the performance of the models with data and present results. Finally in Section \ref{condAndFW} conclusions are presented, along with suggestions for further work.

\section{Hidden Markov Models}
\label{hmmIntroSection}
An HMM is a Bayesian state space model that assumes discrete time unobserved (hidden or latent) states \citep{gales2008application}.  The basic assumptions of a Markov state space model are firstly that states are conditionally independent of all other states given the previous state, and secondly that observations are conditionally independent of all other observations given the state that generated it.

\subsection{Literature Review of HMM in Finance and Economics}
In the 1970's Leonard Baum was one of the first researchers to work with what is now known as an HMM. He applied the methodology to securities' trading for the hedge fund Renaissance Technologies \citep{baum1970maximization, Teitelbaum2008a}.  Since then HMMs have been used extensively in finance and economics \citep{bhar2004hidden, mamon2007hidden}.  The first widely attributed public application of HMM's to finance and economics was by James Hamilton in 1989 \citep{hamilton1989new}.  In his seminal paper, Hamilton views the parameters of an autoregression as the outcome of a discrete Markov process, where the observed variable is GNP and the latent variable is the business cycle. By observing GNP, the position in the business cycle can be estimated and future activity predicted.

Following Hamilton's paper there has been much Bayesian work discussing estimation of these models and providing financial and economic applications, most of which focus on \emph{Markov chain Monte Carlo} (MCMC). MCMC is a means of providing a numerical approximation to the posterior distribution using a set of samples, allowing approximate posterior probabilities and marginal likelihoods to be found.   Two excellent reviews of the field of Bayesian estimation using MCMC are given by Chib \citep{chib2001markov} and Scott \citep{scott2002bayesian}. Noteworthy papers applying Bayesian estimation techniques include; Fruhwirth-Schnatter applies MCMC to a clustering problem from a panel data set of US bank lending data, where model parameters are time-varying \citep{fruhwirth2001markov}. Shephard applies the Metropolis algorithm to a non-Gaussian state space time series model and illustrates the technique by seasonally adjusting a money supply time series \citep{shephard1994partial}.  McCulloch et al apply a Gibbs sampler for parameter estimation in their Markov switching model and illustrate their technique using the growth rates of GNP \citep{mcculloch1994statistical}.  Meligkotsidou et al tackle interest rate forecasting with an non-constant transition matrix using an MCMC reversible jump algorithm for predictive inference \citep{meligkotsidou2011forecasting}.  Less commonly applied in the economic and financial literature is the technique of \emph{variational Bayes} (VB) \citep{attias1999inferring}.   VB provides a parametric approximation to the posterior, often using independence assumptions, in a computationally efficient manner.   McGrory et al apply VB to estimate the number of unknown states along with the model parameters from the daily returns of the S\&P500 \citep{mcgrory2009variational}.  Finally, the debate between learning in HMM's using frequentist methods such as expectation maximization (EM), versus Bayesian methods such as MCMC is reviewed by Ryden who highlights poor mixing and long computation times as potential computational disadvantages of MCMC \citep{ryden2008versus}.

Many different extensions and modifications to the ``vanilla'' HMM have been proposed, and applied to economics and finance. \emph{Input output HMMs} (IOHMM) include inputs and outputs and can be viewed as a directed version of a \emph{Hidden Random Field} \citep{bengio1995input,kakade2002alternate}.  Unlike HMMs, the output and transition distributions are not only conditioned on the current state, but are also conditioned on an observed input value. Bengio et al carry out learning in an IOHMM using a feed-forward neural network. Kim et al use an IOHMM to model stock order flows in the presence of two hidden market states \citep{kim2002modeling}. HMMs are a generalization of a \emph{mixture model} where latent variables control the mixture component to be selected for each observation. In a mixture model, the latents are assumed to be i.i.d. random variables, as opposed to being related through a Markov process as in an HMM \citep{Bishop2006}.  Liesenfeld et al apply a bivariate mixture model to stock price and trading volume \citep{liesenfeld2001generalized}. In their model, the behavior of volatility and volume results from the simultaneous interaction of the number of information arrivals and traders' sensitivity to new information, both of which are treated as latent variables.  In an \emph{hierarchical HMM} (HHMM), each state is itself an HHMM, allowing modelling of ``the complex multi-scale structure which appears in many natural sequences'' \citep{fine1998hierarchical}. Wisebourt et al generate a measure of the limit order book imbalance and uses it to drive latent market regimes inside an HHMM \citep{Wisebourt2011a}.  \emph{Poisson HMMs} (PHMM) are a special case of HMMs where a Poisson process has a rate which varies in association with changes between the different states of a Markov model \citep{scott2002bayesian}. Branger et al apply a PHMM to model jumps in asset price in order to help inform contagion risk and portfolio choice \citep{brangerpartial}.  \emph{Hidden semi-Markov models} (HSMM) have the same structure as a HMM except that the unobservable process is semi-Markov rather than Markov. Here the probability of a change in the hidden state depends on the amount of time that has elapsed since entry into the current state \citep{yu2010hidden}.   Bulla et al apply a HSMM to daily security returns in order to capture the slow decay in the autocorrelation function of the squared returns, which HMMs fail to capture \citep{bulla2006stylized}. Finally, \emph{factorial HMMs} (FHMM) distribute the latent state into multiple state variables in a distributed manner, allowing a single observation to be conditioned on the corresponding latent variables of a set of independent Markov chains, rather than a single Markov chain \citep{ghahramani1997factorial}. Charlot applies an FHMM to design a new multivariate GARCH model with time varying conditional correlation \citep{charlotmodelling}.

Applications of HMMs in finance and economics range extensively, with latent variables including the business cycle \citep{gregoir2000measuring}, inter-equity market correlation \citep{bhar2003new}, bond-credit risk spreads \citep{thomas2002hidden}, inflation \citep{kim1993unobserved, chopin2004bayesian}, credit risk \citep{giampieri2005hidden}, options pricing \citep{buffington2002american}, portfolio allocation \citep{elliott2002portfolio, roman2010hidden}, volatility \citep{rossi2006volatility, dueker1997markov}, interest rates \citep{elliott2007term, ang2002regime}, trend states \citep{dai2010trend, pidan2011selective} and future asset returns \citep{shi1997taking, hassan2005stock, dueker2007can}.

This paper relates to the broader field of research into the prediction of security returns by exploiting the momentum effect. To our knowledge, no other authors have considered momentum as a latent variable in a HMM setting. However Christensen et al have considered a latent momentum formulation in a Bayesian filtering setting \citep{christensen2012forecasting}. In this paper the authors track a continuous latent momentum state of a time series using a Rao-Blackwellized particle filter. The paper finds that the predictions are statistically significant when applied to a portfolio of futures in the presence of transaction costs.  In general terms it is expected that an HMM would be able to outperform the Rao-Blackwellized particle filtering formulation. This is because an HMM with lots of states can model arbitrary transitions between trend states, e.g. a sudden reversal of trend at the top of the market, whereas a linear Gaussian model is limited to linear changes.

\subsection{An HMM for Trading Momentum}
\label{formingTheModel}
\begin{figure}[!ht]
    \begin{center}
        {\includegraphics[width=3.5in,keepaspectratio]{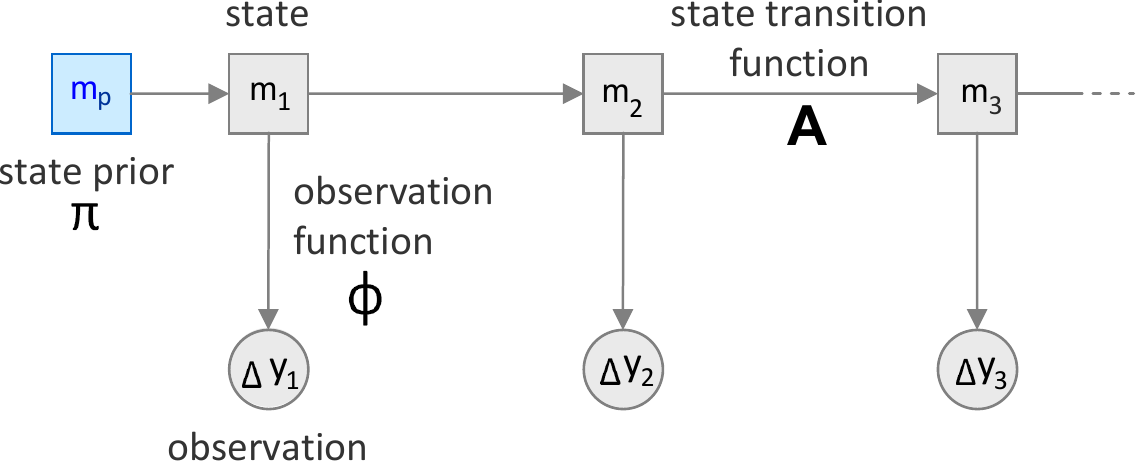}}
        \caption{A state space model for a discretized continuous observed state  $\Delta y$ (the change in price) and a discrete hidden state $m$ (the trend). The relationship between the latent variables and the system parameters is shown.}
        \label{markovM2}
    \end{center}
\end{figure}

Our model is based on the concept of a noisy trend, where the trend is a latent state and the price series is Brownian with a stochastic drift.   In order to forecast the next time step in the HMM we begin with a distribution over the current hidden state and use the transition function to propagate this distribution forward in time. At the next time step we are able to infer the most likely hidden state and generate a predictive distribution over observables. This is done by taking a weighted average of the conditional distribution of the observations where the weights are from the distribution over the hidden state.  We are not interested in predicting price, an arbitrary value, rather the change in price or the \emph{return}. Let $y_t$ be the price, such that ${\bf Y} = \{y_1,\dots, y_T\}$ and $\Delta y_t = \log \nicefrac{y_t}{y_{t-1}}$ be the return, such that $\Delta {\bf Y} = \{ \Delta y_2, \dots, \Delta y_T\}$. In our model $\Delta y_t$ (the observation) is influenced by a hidden, unobserved state, $m_t$ (the trend, where $\nicefrac{d \Delta y}{dt}$ is a noisy estimate of $m_t$), such that ${\bf M} = \{m_1,\dots, m_T\}$.  In order to find $E(\Delta y_{t}| \Delta y_{1:t-1})$, a two step process of learning followed by inference is carried out. This model is shown in Figure \ref{markovM2}.

The intuition behind the model is that security returns can be modelled as a noisy trend process and that while return can be observed, the trend state cannot be and must be inferred.   While the MACD algorithm of Section \ref{mainIntro} attempts to find the true value of this hidden state empirically by use of a digital filter, in this paper we model the observations and the hidden trend state explicitly, therefore allowing interpretation of all the parameters in a meaningful way.  An additional advantage of an HMM formulation over MACD is that HMMs are able to track trends in a much more flexible way, by encoding non-linear relationships between the states. This allows for sudden changes to a new trend, whereas digital filters inevitably have some delay in response depending upon their frequency response.  Finally, we note that digital smoothing filters can often be written equivalently as the stationary solution of particular linear state-space models \citep{harvey1991forecasting}.

Time series, such as security returns, can be \emph{synchronous} or \emph{asynchronous}. A synchronous time series is one where the time stamps lie on a regular grid. The grid spacing is referred to as the sampling frequency.  An asynchronous time series is one where the time stamps do not lie on a regular grid. Raw security returns are generally asynchronous, but are often sampled to make them synchronous. Security returns are not continuous in value, but lie on a discrete price grid, with a grid spacing defined on a security specific basis by the exchange. This grid spacing is called the \emph{tick size}, $\alpha$.  The state space of the latent states consists of a total of $K$ possible values of trend. An upper limit to $K$ can be found by knowing the grid size and calculating $\Omega = \max{\left(|\Delta {\bf Y}| \right)}$.  The latent variables are indexed on this grid as,
\begin{eqnarray}
  m_t  &\in& \left\{1, \dots, k, \dots, K \right\}
  \label{discreteGrid01}
\end{eqnarray}
where $k$ refers to the $k^{\text{th}}$ latent state. By fixing $K$ the set of time-dependent trend terms can be specified by ${\bf M}$.  The observations depend on the latent states according to; return at time $t$ is equal to the trend term, plus Gaussian noise,
\begin{equation}
    \Delta y_t = \mu_{m_t} + \epsilon_t \hspace{1cm} \epsilon_t \sim \text{Norm}(0, \sigma_{m_{t}}^{2})
    \label{PLR}
\end{equation}
Given the indexed grid of Equation \eqref{discreteGrid01}, one mode of initialization would be $\mu_{1:K} = \left\{-\Omega, -(\Omega - \alpha),  -(\Omega - 2\alpha), \dots, 0, \dots, (\Omega - 2\alpha), (\Omega - \alpha), \Omega\right\}$.  The Gaussian assumption of Equation \eqref{PLR} could be replaced with any other parametric distribution (for example, fat-tailed Cauchy) or a non-parametric approach (for example, kernel density estimation).  The resulting conditional distribution from Equation \eqref{PLR} is,
\begin{eqnarray}
    p(\Delta y_t | m_t) = \text{Norm}(\Delta y_t ; \mu_{m_t}, \sigma_{m_{t}}^{2}) \nonumber
\end{eqnarray}
As the latents lie on a discrete grid, yet the noise model is continuous, the implementation is required to discretize the Gaussian noise variable to ensure the results lie on the grid.  The joint distribution of this state space model is therefore given by,
\begin{eqnarray}
 \label{sozc}
    p(\Delta {\bf Y},{\bf M}) = p(m_1) \left[ \prod_{t=2}^{T} p(m_t|m_{t-1}) \right] \prod_{t=1}^{T} p(\Delta y_t|m_t) \nonumber
\end{eqnarray}
Given the HMM and the observations $p({\Delta \bf Y}, {\bf M})$ one can deduce information about the states occupied by the underlying Markov chain $m_{1:T}$.  What we are interested in finding in this model is the probability of a trend given all our observations of price up to now, $p(m_t | \Delta y_{1:t})$, also known as the \emph{filtering distribution}.

\subsection{Model Parameters}
Our model requires that the transition matrix ${\bf A}$, emission matrix ${\boldsymbol \phi}$ and latent node initial value $\pi_1$ are known a-priori. Together these form the parameters of our model $\bf \Theta = \{{\bf A}, \pi, {\boldsymbol \phi}\}$, as shown in Figure \ref{markovM2}. Finding {$\bf \Theta$} constitutes the learning phase of the HMM.   This batch approach to learning suits the structure of the financial markets, as parameter estimation can be done using the previous $H$ days of market data, when the market is shut.  Before discussing learning, the connection between the hidden states ${\bf M}$ and the model parameters ${\bf \Theta}$ is explained. ${\bf A}$ specifies the probability of transitions between the latent states, ${\bf \pi}$ is the probability of the initial latent state and ${\boldsymbol \phi}$ is the probability of the observed return occurring.  The connection between parameter ${\boldsymbol\phi}$ and Equation \eqref{PLR} is that ${\boldsymbol\phi}(\Delta y_t)$ follows a Gaussian distribution.  In this paper four different \emph{off-line} learning approaches are considered,\newline
\begin{enumerate}[noitemsep,nolistsep]
  \item ${\bf \Theta}$ is learnt using piecewise linear regression (PLR).
  \item ${\bf \Theta}$ is learnt using the Baum-Welch algorithm.
  \item ${\bf \Theta}$ is learnt using Markov Chain Monte Carlo (MCMC).
  \item ${\bf \Theta}$ is learnt using the Baum-Welch algorithm in the presence of side information.
\end{enumerate}
PLR and Baum-Welch are both frequentist methods, while MCMC is a Bayesian method. The inference phase of this paper is purely Bayesian.   At this point we consider the correctness of combining frequentist and Bayesian methods in the same model.   The core aim of this paper is to produce the best predictive performance possible, irrespective of methodology used and so from a philosophical viewpoint we are agnostic.  From a practical point of view, frequentist methods are more commonly found in the trading industry. It is reasoned this is due to the relative simplicity of the methods, the parsimony of the models and the associated low computational loads. In particular, trading practitioners tend to dislike complex models due to the risks associated with model failure being low-probability, high-impact. These risks are easier to understand and monitor in simple models.

The major issue when learning is the transient nature of the latent state and how stable its estimated means are. In order to ensure the most accurate estimation possible, the means of the $K$ Gaussian distributions (the trends) are efficiently estimated using short windows of data. This is implemented using a rolling window of data that consists of 23 trading days (one month). This window size approximately agrees with the lowest frequency information we are trying to exploit in our system.

The mixing of frequentist learning with Bayesian inference is a well established approach in the literature, for example Andrieu et al estimate static parameters in non-linear non-Gaussian state space models using EM type algorithms \citep{andrieu2003online}.   Other examples of merging frequentist and Bayesian methodologies are given by Gelman \citep{gelman2011induction} and Jordan \citep{jordan2009you} who suggest using Bayesian inference coupled with frequentist model-checking techniques. Such an approach gives the performance benefits of using a Bayesian prior, while allowing for the easily checkable assumptions given by frequentist confidence intervals. Completely integrated Bayesian techniques to our problem do exist, such as particle MCMC which allows for fully Bayesian learning and inference, however such techniques suffer from unpractically high computational complexity \citep{andrieu2010particle}.

\subsubsection{State Transition Matrix}
A conditional distribution for the latent variables $p(m_t|m_{t-1})$ is specified. Because the latent variables can take one of $K$ values, this distribution is the transition matrix ${\bf A}$, size ($K \times K$). The transition probabilities are given by,
\begin{eqnarray}
{\bf A} =
\left(\begin{matrix}a_{1,1}&a_{1,2}&\dots&a_{1,K}\\
a_{2,1}&a_{2,2}&\dots&a_{2,K}\\
\vdots&\vdots&\ddots&\vdots\\
a_{K^{'},1}&a_{K^{'},2}&\dots&a_{K^{'},K}\\
\end{matrix}\right)
\label{pTransmat} \nonumber
\end{eqnarray}
where 
\begin{eqnarray}
    a_{k^{'},k} = p(m_{t} = M_{k} | m_{t-1} = M_{k^{'}}) ~~   k^{'},k = 1, \dots, K \nonumber
\end{eqnarray}
i.e. the probability of making a particular transition from state $k^{'}$ to state $k$ in one time step is given by $a_{k^{'},k}$. The diagonal corresponds to the probability of the system staying in its current state, while the lower diagonal corresponds to the system moving to a negative price trend and the upper diagonal corresponds to moving to a positive price trend. ${\bf A}$ has $K(K-1)$ independent parameters and each row of ${\bf A}$ is a probability distribution function such that $\sum_k a_{k^{'}k} =1$.

\subsubsection{Emission Matrix}
The probability of an observation given the hidden state is given by the emission matrix ${\boldsymbol \phi}$.  This matrix is a set of parameters governing the conditional distribution of the observed variables $p({\Delta \bf Y} | {\bf M}, {\boldsymbol \phi})= {\boldsymbol \phi}_k(\Delta y)$.  In a discrete HMM model, an emission matrix is output of size the number of states in the hidden representation by the number of possible output states.   For our continuous model each one of the $K$ states has an associated output distribution of a single univariate discretized Gaussian, with a mean $\mu_k$ and a variance $\sigma_{k}^2$, as given by Equation \eqref{emissionMatrix2}.
\begin{eqnarray}
\phi_k({\Delta y}) &\propto& \text{Norm}({\Delta y} ; \mu_k, \sigma_k^2 ) \nonumber\\
                            &=& \frac{\text{Norm}({\Delta y} ; \mu_k, \sigma_k^2 )}{ \sum_{\Delta y \in \cal{Y}} \text{Norm}({\Delta y} ; \mu_k, \sigma_k^2 )}
\label{emissionMatrix2}
\end{eqnarray}
where $\cal{Y}$ denotes the set of all possible $\Delta y$.

\subsubsection{Initial Latent Node}
The initial latent node $m_1$ is special in that it does not have a parent node and so it has a marginal distribution $p(m_1)$ represented by a vector of probabilities ${\bf \pi}$ with elements $ \pi_k \equiv p(m_{1} = k)$.

\subsubsection{Number of Unknown States}
As the number of latent momentum states $K$ is unknown, estimating $K$ is a \emph{model selection} problem.  There are various methodologies for determining $K$, both heuristic and formal, frequentist and Bayesian. We summarize some of these techniques here,
\begin{itemize}[noitemsep,nolistsep]
  \item Cross validation \citep{kohavi1995study}. Segment the data set into training and test portions. Select $K$ which gives the best predictive performance on the training data set and then apply it to the test data set.
  \item Generalized likelihood ratio tests \citep{vuong1989likelihood}. The ratio of two model's likelihoods is used to compute a p-value, which allows the null hypothesis to be accepted or rejected.
  \item Penalized likelihood criteria, such as Bayesian information criterion (BIC) \citep{schwarz1978estimating} and Akaike information criterion (AIC) \citep{akaike1974new}. These criteria penalize the maximized likelihood function by the number of model parameters. The disadvantage, is that they do not provide any measure of confidence in the selected model.
  \item Approximate Bayesian computation \citep{toni2009approximate}. Simulation based model selection.
  \item Bayesian model comparison  \citep{kass1995bayes}. Theoretically powerful, but difficult to apply in practice. This approach is often approximated by MCMC \citep{gilks1996markov}.
\end{itemize}
The posterior probability $p({\cal M}_k | {\Delta {\bf Y}}, {\bf \Theta})$ of a model ${\cal M}_k$ given data ${\Delta {\bf Y}}$ is given by Bayes theorem,
\begin{eqnarray}
  p({\cal M}_k | {\Delta {\bf Y}}, {\bf \Theta}) &=& \frac{p({\Delta {\bf Y}} | {\cal M}_k, {\bf \Theta}) p({\cal M}_k)}{p({\Delta {\bf Y}})} \nonumber
\end{eqnarray}
For two different models ${\cal M}_1, {\cal M}_2$ with parameters $\Theta_1, \Theta_2$, the \emph{Bayes factor} ${\cal B}$ can be used to carry out model selection,
\begin{eqnarray}
  {\cal B} = \frac{p({\Delta {\bf Y}} | {\cal M}_1)}{p({\Delta {\bf Y}} | {\cal M}_2)} = \frac{\int p(\Theta_1 | {\cal M}_1) p({\Delta {\bf Y}} | \Theta_1, {\cal M}_1)d\Theta_1}{\int p(\Theta_2 | {\cal M}_2) p({\Delta {\bf Y}} | \Theta_2, {\cal M}_2)d\Theta_2}  \nonumber
\end{eqnarray}
The chosen model is simply the model with the highest integrated likelihood $p({\cal M}_k | {\Delta {\bf Y}}, {\bf \Theta})$. However, at times the prior $p({\bf \Theta}|{\cal M}_k)$ is unknown and so the logarithm of the integrated likelihood can be approximated by the BIC. More accurate selection of $K$ requires evaluating the marginal likelihood $\int p({\bf \Theta} | {\cal M}_k) p({\Delta {\bf Y}} | {\bf \Theta}, {\cal M}_k)d{\bf \Theta}$, however this is an extremely difficult integral to calculate. Dealing with this integral is covered in Section \ref{mcmcSection} once MCMC has been introduced. In the following Section, each method of learning uses one of the above techniques to estimate $K$.

\section{Learning Phase}
\label{learningSection}
The three independent methods of learning ${\bf \Theta}$ are now presented. Results are shown for applying the methods to one year's worth of data at one minute sampling frequency from the ES future, a traded security. Full details of the dataset and and its processing are described in Section \ref{dataSimSection}.

\subsection{Piecewise Linear Regression}
\label{defaultLearningA}
The ``default case'' is presented as a baseline against which other methods of learning can be compared. ${\bf A}$ is initialized as,
\begin{eqnarray}
        a_{k^{'},k} =
                \begin{dcases*}
                        \beta,                     & $k^{'}=k$       \\
                        \nicefrac{1-\beta}{K-1},   & $k^{'} \neq k$
                \end{dcases*}
                \label{defaultA}\nonumber
\end{eqnarray}
where $\beta$ is the probability of the state staying in its current state and is set as $\beta=0.5$.   The $\nicefrac{1-\beta}{K-1}$ term reflects that fact that in the absence of conditioning information, no state is more likely than any other state, though it is most likely to stay in its current state and thus is described as ``sticky''.

\emph{Change points} ${\bf P}$ in the price ${\bf Y}$ represent breaks between latent momentum states (i.e. trends).  Using \emph{piecewise linear regression} (PLR) on the training data set, ${\bf P}$ is found \citep{Oh2011a}. PLR gives two things - firstly a state sequence which can be used in learning later and secondly, the model mean $\mu_k$ and variance $\sigma_k^2$. PLR is simply ordinary least squares carried out over segmented data, with change points tested for by t-stats. For each segment of data that contains a trend, $\mu_k$ is the gradient of the regression and $\sigma_k^2$ the variance, found from the maximum likelihood estimate for a Gaussian noise model,
\begin{eqnarray}
    \sigma  = \sqrt{\frac{\sum_{t=1}^{T} \epsilon_{t}^2}{T}}\nonumber
    \label{residuals}
\end{eqnarray}
where $\epsilon$ are the regression residuals. The presence of autocorrelation would suggest that PLR was not working correctly and so is checked for using the Durbin-Watson test \citep{durbin1971testing}.   Finally, it is noted that many other approaches exist for change point detection, for example \citep{Adams2007a, punskaya2002bayesian}.

For the default case, the number of hidden states is found using cross validation and is set to $K=2$.

\subsection{Baum-Welch}
\label{baumWelch}
The Baum-Welch algorithm is a special case of the EM algorithm which can be used to determine parameter estimates in an HMM when the state sequence of the latents is unknown \citep{baum1970maximization}.   The algorithm attempts to find the sequence of latent states ${\bf M}$ which will maximize the likelihood of having generated ${\Delta \bf Y}$ given ${\bf \Theta}$,
\begin{eqnarray}
{\hat{\bf \Theta}} =  \underset{\bf \Theta}{\operatorname{argmax}} ~ p({\Delta \bf Y} | {\bf \Theta})\nonumber
\end{eqnarray}
Finding this global maximum is intractable as it requires enumerating over all parameter values ${\bf \Theta}_k$ and then calculating $p({\Delta \bf Y} | {\bf \Theta}_k)$ for each $k$.   Baum-Welch avoids this global maximum and instead settles for a local maximum.  As with other members of the EM class, this is achieved by computing the expected log-likelihood of the data under the current parameters and then using this to iteratively re-estimate the parameters until convergence.

In the first step of the algorithm (the \emph{E-step}), Baum-Welch uses the forward-backward algorithm, which finds the smoothing distribution $p(k | \Delta y_{1:T})$.  The Forward algorithm gives $\alpha_t(k)$, which is the probability that the model is in state $k$ at time $t$, based on the current parameters. The Backward algorithm gives $\beta_{t+1}(k)$ which is the probability of emitting the rest of the sequence if we are in state $k$ at time $t+1$, based on the current parameters. For large numbers of observations, numerical under-flow can occur, hence in implementation log-probabilities are used \citep{kingsbury1971digital}. 

The second step of the algorithm (the \emph{M-step}), sees successive iterations of the algorithm update ${\bf \Theta}$ improving the likelihood up to some local maximum. This is done by calculating the \emph{occupation probabilities} $\gamma_t(k)$ which is the probability of the model occupying state $k$ at time $t$.  These probabilities are then used to find the maximum likelihood estimates of ${\bf A}$ and ${\boldsymbol \phi}$ \citep{juang1986maximum}.

\begin{figure}[!ht]
    \begin{center}
        {\includegraphics[width=3in,keepaspectratio]{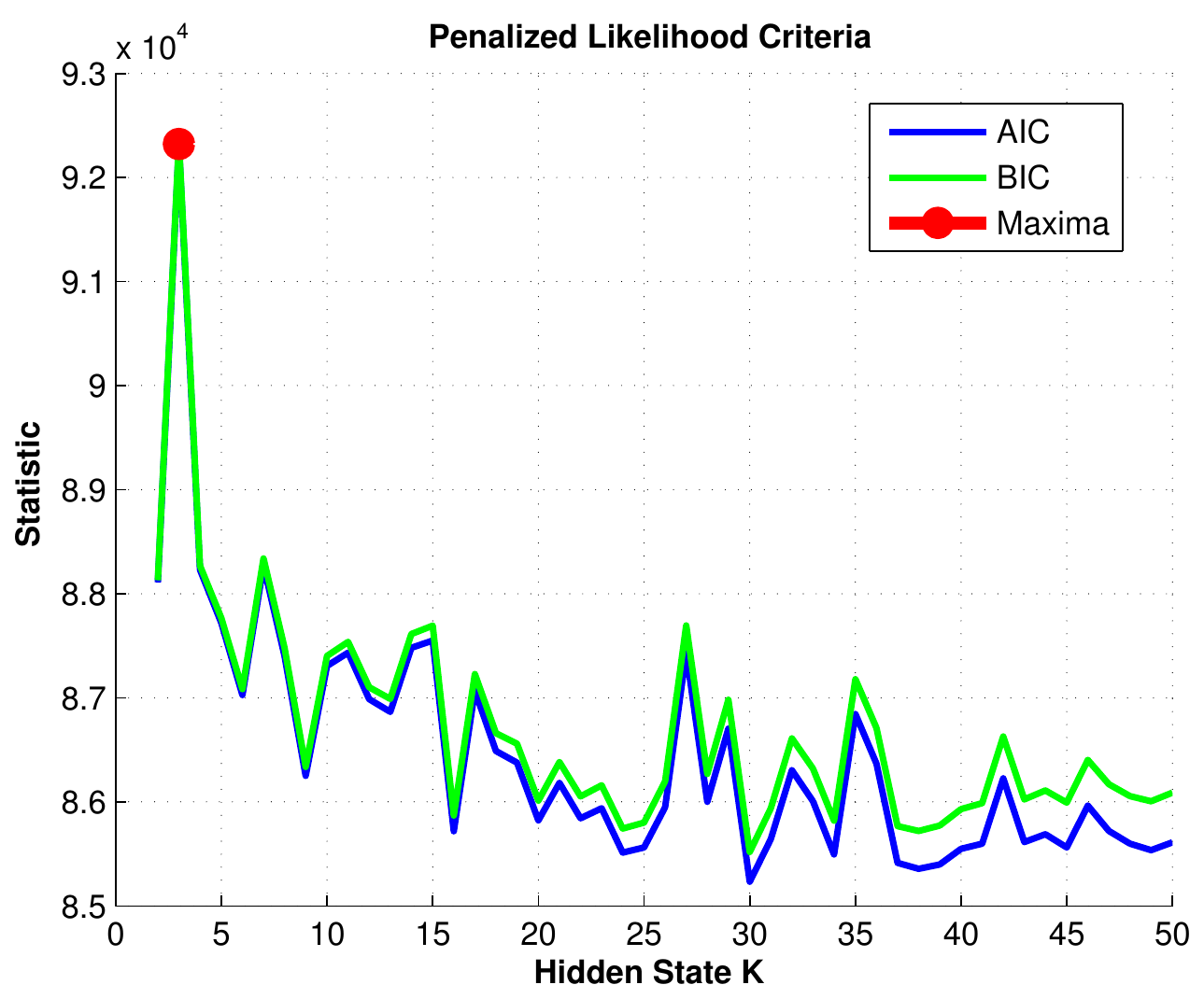}}
        \caption{Penalized likelihood criteria. Finding the number of hidden states using Baum-Welch. The optimal model of $K=3$ is shown by a red dot.}
        \label{baumWelchFindK}
    \end{center}
\end{figure}

Baum-Welch is used to find $K$ by maximizing the log-likelihood of $k=1,\dots,50$ models. Penalized likelihood criteria are calculated for each model and the maximum value $K=3$ selected. The results are shown in Figure \ref{baumWelchFindK}.\newline
Baum-Welch requires estimates for initial value of the emission and transition matrices.  A ``flat start model'' is defined by setting all the values of ${\bf A}$ to be equal and ${\boldsymbol \phi}$ to the global mean/variance of the data. The problem with this approach is that, depending on how the initial HMM parameters are chosen, the local maximum to which Baum-Welch converges to may not be the global maximum.   Convergence is deemed to have occurred when either a certain number of iterations have passed or a certain log-likelihood tolerance has been met.  In order to hit the global maximum, \emph{good initialization is crucial}. To avoid local minima, a prior is set over ${\bf \Theta}$ using training data ${\bf Z}$.  Applying Baum-Welch to ${\bf Z}$ it is noted that the square root of the model variances $\sigma_k^2$ is of the same order of magnitude as the tick-size $\alpha$ for the ES contract.  This is as expected as the algorithm is unable to predict with an accuracy smaller than the grid size.   Learning the covariance structure (untied) can result in a implementation issue that for one or more states the local maxima might settle on a small number of data points, giving $\sigma_k^2 \rightarrow 0$, preventing the log-likelihood increasing at each iteration of the M-Step.  This is dealt with in our implementation by never allowing the variance to decrease below a fraction of the tick-size, $\nicefrac{\alpha^2}{2}$.  The Baum-Welch algorithm is shown in Algorithm \ref{skynetBW}.  The notation of $k$ to refers to a particular state and not the indicator variable $m_t$, as that is path-dependent.
\begin{algorithm}
    \caption{HMM Baum-Welch. \hspace{0.2cm} \\ ${\bf \hat{\Theta}} = \text{BW}({\bf Z}, K)$}
    \begin{algorithmic}[1]
    \label{skynetBW}
        \STATE{\bf Initialize}
        \STATE ${\bf \Theta}\left\{{\bf A}, {\boldsymbol \phi}\right\} = \text{extract}({\bf Z})$          \COMMENT{Extract initial parameters from the estimate}
        \WHILE{$q < maxIterations$}    \STATE{Go around loop until parameters converge or $tol$ is met}
            \STATE{\bf Forward Pass}
            \STATE $\alpha_1(k) = p(m_1)p(z_1|m_1)$ \COMMENT{Initialization}
            \FOR{$t=2$ to $T$}
                \STATE $\alpha_t(k) = \sum_{m_{t-1}}p(z_t|m_t)p(m_t|m_{t-1})\alpha_{t-1}(k)$   \COMMENT{Generate a forwards factor by eliminating $m_{t-1}$}
            \ENDFOR
            \STATE{\bf Backward Pass}
            \STATE $\beta_t(k) =1$ \COMMENT{Initialization}
            \FOR{$t=T-1$ to $1$}
                \STATE $\beta_t(k)=\sum_{m_{t+1}} p(z_{t+1} | m_{t+1}) p(m_{t+1} | m_t) \beta_{t+1}(k)$   \COMMENT{Generate a backwards factor by eliminating $m_{t+1}$}
            \ENDFOR
            \STATE{\bf Occupation Probabilities}
            \STATE $\gamma_t(k) = \frac{\alpha_t(k) \beta_t(k)}{p({z_t})}$
            \STATE{\bf Parameter Estimation}
            \STATE ${\bf \mu}(k) = \frac{\sum_{t=1}^{T} \gamma_t(k) z_t}{ \sum_{t=1}^{T} \gamma_t(k) }$
            \STATE $\sigma^2(k) = \frac{\sum_{t=1}^{T} \gamma_t(k)(z_t - \mu_{k})(z_t - \mu_{k})^T}{\sum_{t=1}^{T} \gamma_t(k) }$ \COMMENT{Marginalizing over $k$ gives ``tied'' $\sigma^2$}
            \STATE ${\boldsymbol \phi} \sim \text{Norm}\left({\bf Z} ; \mu_k ,  \sigma_k^2 \right)$
            \STATE ${\bf A} = \frac{1}{p(z)}\frac{\sum_{t=1}^{T} \alpha_t(k) a_{k^{'},k} \phi_k(z_{t+1}) \beta_{t+1}(k)z_t   }{\sum_{t=1}^{T}\gamma_t(k)}$
            \STATE $\text{score} = p({\bf Z} | {\bf A}, {\boldsymbol \phi})$
            \STATE{\bf Terminate}
            \IF{score < tol}
                \STATE ${\bf \hat{\Theta}} = \left\{{\bf A}, {\boldsymbol \phi} \right\}$ \COMMENT{Maximum likelihood estimates}
                \STATE $\text{return}({\bf \hat{\Theta}})$
            \ENDIF
        \ENDWHILE
    \end{algorithmic}
\end{algorithm}

\subsection{Markov Chain Monte Carlo}
\label{mcmcSection}
MCMC methods are a class of algorithms for sampling from probability distributions based on constructing a Markov chain that has the desired distribution as its equilibrium distribution. By constructing Markov chains for sampling specific densities, marginal densities, posterior expectations and evidence can be calculated.  The Metropolis-Hastings algorithm (MHA) is a simple and widely applicable MCMC algorithm that uses proposal distributions to explore the target distribution \citep{metropolis1953equation}.   MHA constructs a Markov chain by proposing a value for ${\bf \Theta}$ from the proposal distribution, and then either accepting or rejecting this value (with a certain probability).  Given the well established literature on MHA in the financial field, the reader is referred to the review at \citep{chib2001markov}.

In order to find the unknown number of states in a Bayesian framework, a prior distribution is placed on model ${\cal M}_k$ and then posterior distribution of ${\cal M}_k$ is estimated given data ${\Delta {\bf Y}}$,
\begin{eqnarray}
    p({\cal M}_k | {\Delta {\bf Y}}) \propto p({\Delta {\bf Y}} | {\cal M}_k)\times p({\cal M}_k) \nonumber
\end{eqnarray}
where $p({\cal M}_k)$ is the prior, $p({\cal M}_k | {\Delta {\bf Y}})$ is the posterior and the quantity we wish to estimate is the marginalized likelihood $p({\Delta {\bf Y}} | {\cal M}_k)$.  However, as marginal likelihood integration is intractable, simulation based approaches must be used.  There are many ways to approximate this marginal likelihood using MCMC draws, typically done using MHA for each ${\cal M}_k$ separately. However, all known estimators have been shown to be biased \citep{robert2008some}. Another technique from the literature is reversible-jump MCMC (RJMCMC), however this is highly computationally intensive \citep{green1995reversible}.  Based on the lower run-time, $K$ is estimated using marginal likelihoods.  To avoid a biased estimator, this is done using a simulation based approximation of the marginal likelihood called \emph{bridge sampling} \citep{fruhwirth2006finite}. Bridge sampling takes an i.i.d. sample from an importance density and combines it with the MCMC draws from the posterior density in an appropriate way. With bridge sampling, $p({\Delta {\bf Y}} | {\cal M}_k)$ is approximated by,
\begin{eqnarray}
    \hat{p}({\Delta {\bf Y}} | {\cal M}_k) = \frac{L^{-1}\sum_{l=1}^{L} \kappa(\tilde{\theta}^{[l;k]}) p^{*}(\tilde{\theta}^{[l;k]} | {\Delta{\bf Y}},{\cal M}_k)} {N^{-1}\sum_{n=1}^{N}\kappa(\breve{\theta}^{[n;k]})q^{(\breve{\theta}^{[n;k]})}}  \nonumber
\end{eqnarray} 
where $p^{*}(\theta | {\Delta {\bf Y}}, {\cal M}_k) = p({\Delta {\bf Y}} | \theta, {\cal M}_k)\times p(\theta | {\cal M}_k)$, and is the unnormalised posterior density of $\theta$ on ${\bf \Theta}_{k}$, $\kappa$ is an arbitrary function on ${\bf \Theta}_{k}$, $q$ is an arbitrary probability density on ${\bf \Theta}_{k}$, $\breve{\theta}^{[n;k]}$ are samples from the posterior $p(\theta | {\Delta {\bf Y}}, {\cal M}_k)$ obtained using MHA and $\tilde{\theta}^{[l;k]}$ are i.i.d. samples from $q$ \citep{ryden2008versus}.  A drawback to the bridge-sampling approach is that if the number of hidden states is suspected to be larger than about six, then empirically the technique becomes inaccurate and a trans-dimensional approach such as RJMCMC has to be used. This is because it is essential that all modes of the posterior density are covered by the importance density $q(\theta)$, to avoid any instability in the estimators \citep{fruhwirth2006finite}.

\begin{figure}[!ht]
    \begin{center}
        {\includegraphics[width=3in,keepaspectratio]{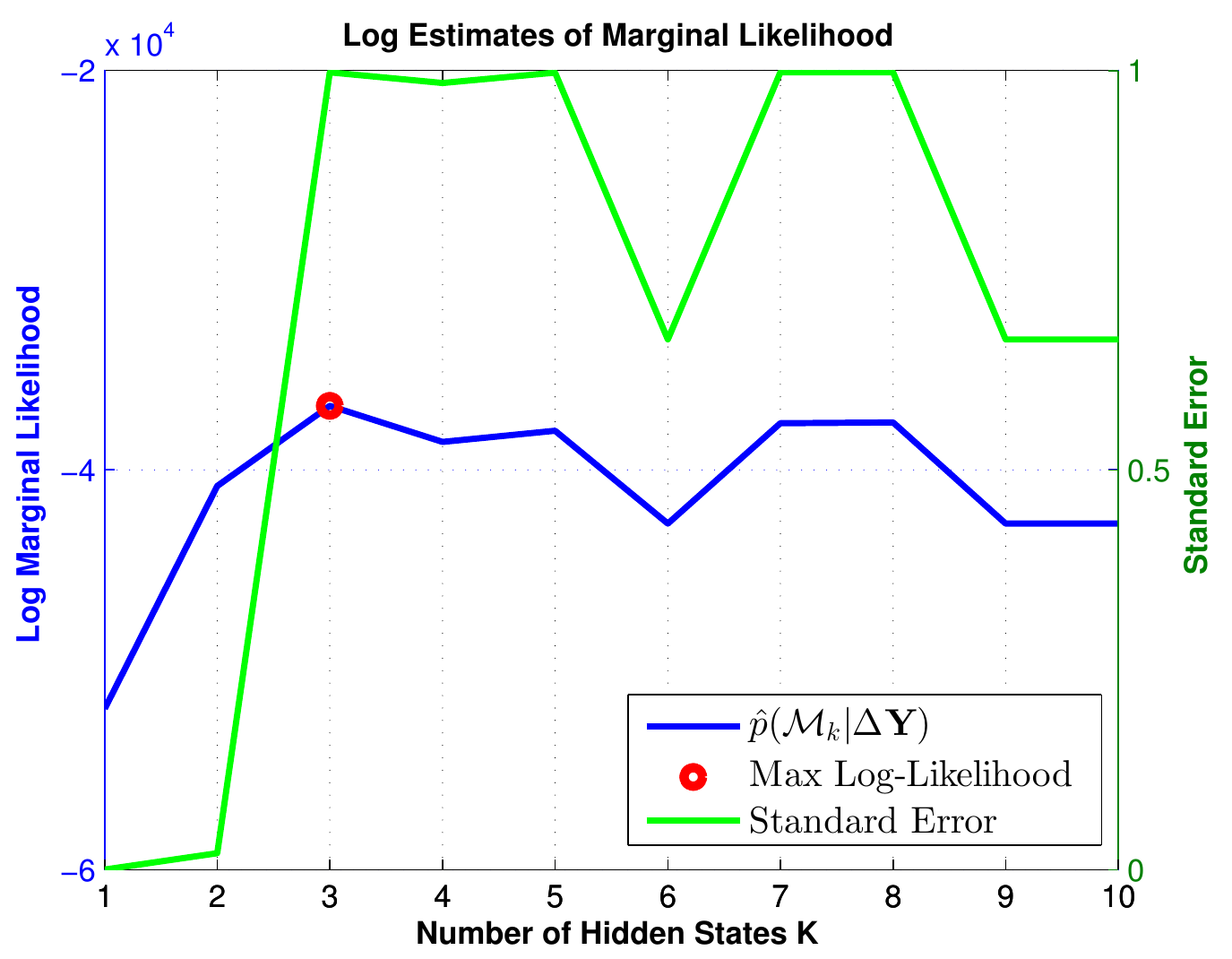}}
        \caption{Log of the bridge sampling estimator of the marginal likelihood $\hat{p}({\Delta {\bf Y}}|{\cal M}_k)$ under the default prior for $K = 1,\dots,10$. The maximum is at $K =3$. On the right-hand axis the standard error is shown for each model.}
        \label{mcmc_marginalLL}
    \end{center}
\end{figure}

A literature review was conducted on the estimation of the number of hidden states in S\&P500 daily return data. Assorted techniques including VB, RJMCMC, EM, penalized likelihood criteria and bridge-sampling all estimated the data to contain between 2 and 3 hidden states \citep{mcgrory2009variational, robert2000bayesian, ryden1998stylized, fruhwirth2008comment, ryden2008versus}.   As a result of this we believe that $K \le 10$, while noting our data sampling frequency is significantly different from that used in the literature (one minute versus daily).   In order to find $K$, a series of mixture distributions of a univariate normal are specified. For each of $k=1,\dots,10$ models the log of the bridge sampling estimator of the marginal likelihood $\hat{p}({\Delta {\bf Y}} | {\cal M}_k)$ is found. The results are shown in Figure \ref{mcmc_marginalLL}.\newline

It can be seen that the largest marginal likelihood is a mixture of three normal distributions, meaning $K=3$ is the number of hidden states suggested by MCMC. Using this number of hidden states, ${\bf \Theta}$ is found. The choice of prior is a critical step in the MCMC process and can lead to significant variations in the posterior probabilities.  A proper prior is defined based on the methodology suggested by $\text{Fr$\ddot{\text{u}}$hwirth-Schnatter}$ \citep{fruhwirth2008comment}.   The prior combines the hierarchal prior for state specific variances $\sigma_k^2$ with a informative prior on the transition matrix ${\bf A}$ by assuming that each row $(a_{i1},\dots,a_{iK})$, $i=1,\dots,K$ follows a Dirichlet $\text{Dir}(e_{i1},\dots,e_{iK})$ prior where $e_{ij}=4$ and $e_{ij} = \nicefrac{1}{(d-1)}$ for $i\ne j$. By choosing $e_{ii} > e_{ij}$ the HMM is bounded away from a finite mixture model \citep{fruhwirth2008comment}.  The vector ${\bf \pi} = \{\pi_1,\dots,\pi_K\}$ of the initial states is drawn from the ergodic distribution of the hidden Markov chain.

As a point estimate is required for ${\bf \Theta}$, we must move from the distributional estimate to a point estimate. This is done by approximating the posterior mode.  The posterior mode is the value of ${\bf \Theta}$ which maximizes the non-normalized mixture posterior density $\log p^{*}({\bf \Theta} | {\Delta {\bf Y}} ) = \log p({\Delta {\bf Y}} | {\bf \Theta} ) + \log p({\bf \Theta})$. The posterior mode estimator is the optimal estimator with respect to the $\nicefrac{0}{1}$ loss function. The estimator is approximated by the MCMC draw with the largest value of $\log p^{*}({\bf \Theta} | {\Delta {\bf Y}} )$.

\begin{figure}[!ht]
    \begin{center}
        {\includegraphics[width=3in,keepaspectratio]{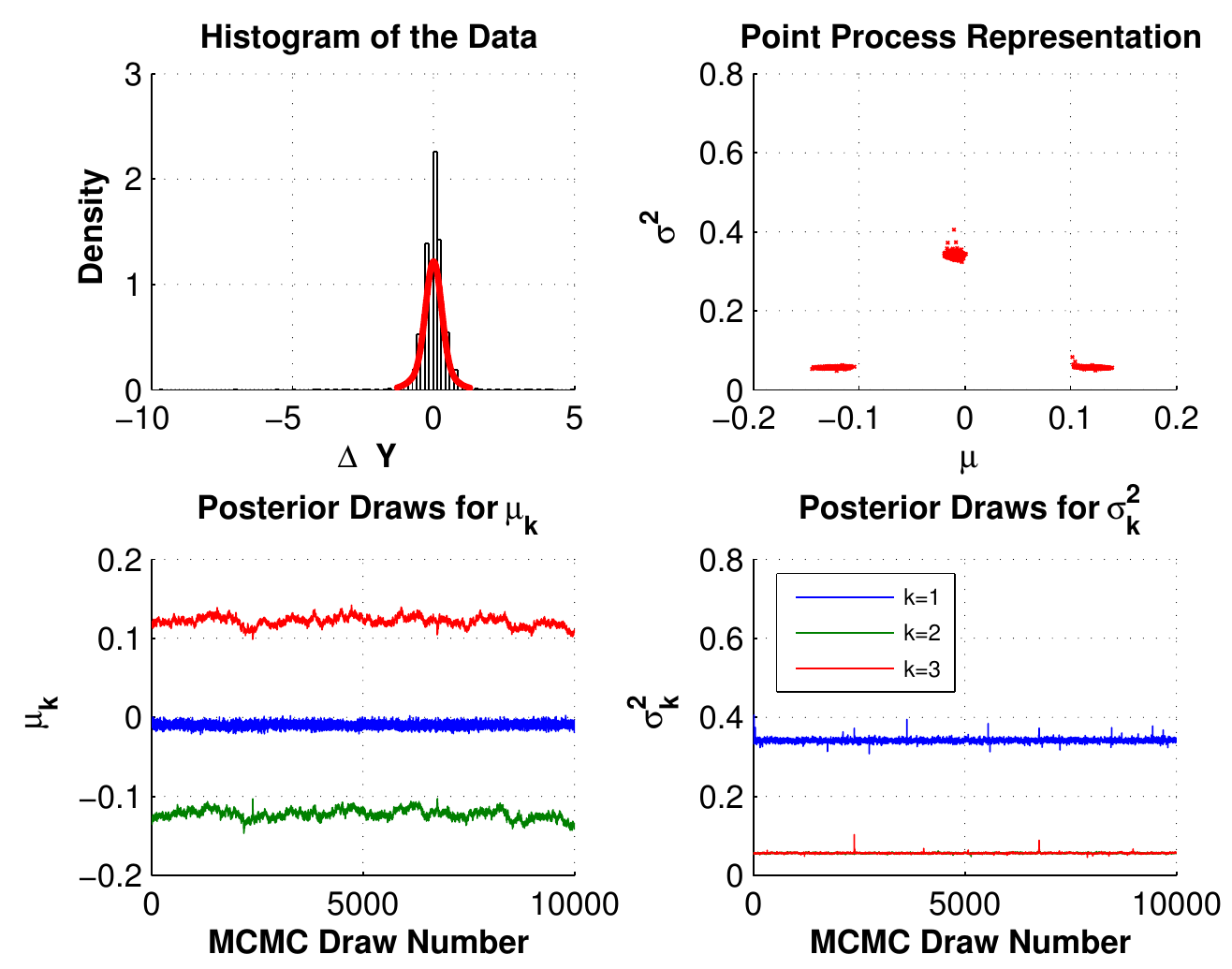}}
        \caption{Markov Chain Monte Carlo by the Metropolis-Hastings algorithm. Subplot one, histogram of the data in comparison to the fitted 3 component Gaussian mixture distribution.  Subplot two, a point process representation for $K=3$.  Subplot three, MCMC posterior draws for $\mu_k$.  Subplot four, MCMC posterior draws for $\sigma_k^2$}
        \label{mcmc_Results}
    \end{center}
\end{figure}

As samples from the beginning of the chain may not accurately represent the desired distribution a ``burn-in'' period of 2,000 draws was used. Run length was set to 10,000 draws. Implementation used the \emph{Bayesf} toolbox with full details of the approach followed found in subsection 11.3.3 of \citep{fruhwirth2006finite}. A selection of the MCMC outputs are shown in Figure \ref{mcmc_Results}.\newline

\subsection{Learning Summary}
In this subsection the major differences the three methods of learning are considered and the results compared. The three techniques for estimating the number of hidden states all gave very similar results. Cross validation for PLR gave $K=2$, penalized likelihood criteria for Baum-Welch gave $K=3$ and bridge-sampling for MCMC gave $K=3$.   For a momentum model both $K=2$ and $K=3$ makes sense, as $K=2$ could correspond to an upward/downward-trending momentum states, with $K=3$ meaning an additional no-trending momentum state. Any higher values of $K$ may just be considered noise. Subplot three of Figure \ref{mcmc_Results} supports this hypothesis by showing that the gradient of the trend is either positive, negative or zero, corresponding to upward/downward/no-trending states.  These observations translate into different conditional means for the two/three normal distributions and are reported in the results Section \ref{dataSimSection}.  The framework or two or three states is appealing as experiments with MACD momentum models have shown only the sign of the predictive signal has traction against the sign of future returns. The magnitude of the signal does not seem able to predict the magnitude of future returns.

The inclusion of the PLR learning allows a ``naive'' estimate of the system parameters to be compared to the formal EM and MCMC techniques.  Both deterministic Baum-Welch and stochastic MCMC use statistical inference to find the number of hidden states and system parameters in an HMM.  Baum-Welch can be used for maximum likelihood inference or for maximum a-posteriori (MAP) estimates.  A Bayesian approach retains distributional information about the unknown parameters which MCMC can be used to approximate.  Baum-Welch computes point estimates (modes) of the posterior distribution of parameters, while MCMC generates distributional outputs.  Both learning approaches have their advantages and disadvantages. One pass of the EM algorithm is computationally similar to one sweep of MCMC, however typically many more MCMC sweeps are run than EM iterations, meaning the computational cost for MCMC is much higher.  Baum-Welch does not always converge on the global maxima, whereas MCMC suffers from the difficulty of choosing a good prior and potentially poor mixing of MCMC.  For MCMC, estimating the number of latent states by a mixture likelihood may be a fragile process. It will obviously depend upon the distributions chosen. If a non-Gaussian distribution were selected, the mixture might be of lower order. This point also applies to the other learning approaches as well.  In summary EM is found to be the simplest and quickest solution \citep{ryden2008versus}.   The relative predictive performance of the three sets of ${\bf \Theta}$ is presented in Section \ref{dataSimSection}.

So far, we have considered the relatively simply specification of two and three state Markov regime switching between Gaussian distributions. This approach is well known to be able to capture some aspects of the nonlinearity of price formation, however it does suffer from overfitting and unobservability in the underlying states.    Chen et al provide an interesting critique of other such approaches applied to forecasting electricity prices \citep{chen2014forecasting}. The authors conclude that a finite mixture approach to regime switching performs best in out-of-sample testing, a methodology that we may look to in future work. In the following section, the sophistication of the model is increased by the inclusion of exogenous information.

\section{Side Information}
\label{sideInfoSection}
In this Section, the case where the probability of any given state in ${\bf A}$ is affected by side information from outside the model is considered. This is important as ${\bf A}$ governs the dynamics of ${\bf Y}$.  In ``classical'' trading models, the $t+1$ return of a security is forecast by a ``signal'' which is a univariate time series, typically synchronous and continuous between $\pm 1$. When this signal is $>0$ the trader will go ``long'' and when the signal is at $<0$ the trader will go ``short''. In the simple case of a portfolio consisting of only one security, the number of lots of security to be traded is directly proportional to the product of the signal magnitude and available capital.    Historically such predictive trading signals are generated from either ``technicals'' or ``fundamentals''. Technicals are signals based on the prior behaviour of the security \citep{Schwagertech1995}. Fundamentals are signals based on upon extrinsic factors (such as economic data) \citep{Schwager1995a}.  In this section two predictive signals are generated and shown to have statistical traction against security returns. In Section \ref{learningWithSideInfo}, the information held in these signals is used when learning ${\bf A}$.  This methodology is quite general and as such could be applied to any technical or fundamental predictor.

Momentum traders often want to combine their momentum signal with one or more extrinsic predictive signals to give a single forecast. This is called the \emph{signal combination problem} for which a variety of different solutions exist, for example, Bayesian model averaging \citep{hoeting1999bayesian}, frequentist model averaging \citep{Wang2009a}, expert tracking \citep{Lugosi2006a} and filtering \citep{Genasay2001a}.  It is noted that our approach of biasing the transition dynamics of an HMM momentum trading system using external predictors seems to be another possible solution to this problem. 

\subsection{Forecasting with Splines}
Splines are now introduced as the methodology by which we condition learning of the transition matrix.  Splines are a way of estimating a noisy relationship between dependent and independent variables, while allowing for subsequent interpolation and evaluation \citep{Reinsch1967a}.    Splines have been used extensively in the financial trading literature, in areas as diverse as volatility estimation \citep{audrino2009splines}, yield curve modelling \citep{bowsher2008dynamics} and returns forecasting \citep{DABLEMONT2010a}.  We use a B-spline as a way of capturing a stationary, non-linear relationship between predictor and security return.  Splines are implemented in MATLAB using the \emph{shape modelling language} toolbox \citep{J.DErrico2011} and the \emph{curve fitting toolbox} \citep{matlabSplines}.  In our experience fitting splines seems to be as much an art as a science with sources of variability including how to treat end-points and the number of knots depending on the degree of ``belief'' in the underlying economic argument of the relationship.  Each spline is forced to be zero mean by setting the integral of the spline to be zero, as the mean value of a function is the integral of that function divided by the length of the support of that function. A zero mean spline ensures that no persistent bias is allowed over the interval of estimation.  For a predictor ${\bf X} = \left\{x_1,x_2,\dots, x_T \right\}$ the learning and subsequent forecasting procedure is shown in Algorithm \ref{splineTrading}.\newline
\begin{algorithm}
    \caption{Learning and Forecasting With Splines.  \hspace{0.2cm} \\ ${\Delta \bf{\hat Y}} = \text{LAFWS}({\bf Y}, {\bf X})$}
    \begin{algorithmic}[1]
    \label{splineTrading}
    \FOR{$n=1$ to $N$}
        \FOR{$t=1$ to $T$}
            \STATE $\Delta y = \log{\left(\frac{y_{nt}}{y_{nt-1}}\right)}$                                        \hspace{2.7cm}    \COMMENT{Take returns}
        \ENDFOR
        \STATE ${\Delta \bar{y}} = \frac{\Delta y - \mu_{\Delta y}}{\sigma_{\Delta y}}$   \hspace{2.5cm}    \COMMENT{Normalize the return, $\Delta \bar{y} \sim \text{Norm}(0, 1^2)$}
        \STATE ${\cal G}_{n-n^{'}:n} = \text{spline}(x_{n-n^{'}:n}, \Delta {\bar{y}_{n-n^{'}:n}})$       \hspace{0.5cm}    \COMMENT{Generate spline ${\cal G}$}
        \IF{$n > n^{''}$}
            \FOR{$t=1$ to $T$}
                \STATE ${\Delta \hat{y}_t} = {\cal G}_{n-n^{''}:n}\left(x_t \right)$                                \hspace{3.0cm}    \COMMENT{Evaluate the spline}
            \ENDFOR
        \ENDIF
    \ENDFOR
    \end{algorithmic}
\end{algorithm}
where $t =1,\dots, T$ is intra-day time and $n=1,\dots, n^{'}, n^{''}, \dots, N$ is inter-day time. Spline evaluation is intra-day while spline learning is inter-day where the spline is ``grown'' over time, allowing it to capture new information and forget old information.  The normalization step for price is carried out using an exponentially weighted moving average process for both mean $\mu_{\Delta y}$ and volatility $\sigma_{\Delta y}$ \citep{pesaran2009model}.   In our code $n^{''}=66$ days with $N= 258$ days and $T=856$ observations per day. In this way the spline is estimated using the previous 66 trading days worth of data, on a rolling basis.

In the next two sections we implement two popular ``off the shelf'' predictors from the literature which exploit intraday effects and use them to generate ${\bf X}$.

\subsection{Predictor I: Volatility Ratio}
\label{volRatioSection}
An extensive body of empirical research exists showing that realized volatility has predictive power against security returns \citep{christoffersen2003a, Hibbert2008a, giot2005a, Burghardt2008a}.   These observations can be explained by showing that the sign dynamics of security returns are driven by volatility dynamics \citep{Kinlay2006a}. Modelling the returns process $\Delta y_t$ as Gaussian with mean $\mu$ and conditional volatility $\sigma_t$ allows for probability distribution function $f$ and a cumulative distribution function $F$.  The probability of a positive return $p(\Delta y_{t+1}) >0$ is given by $F=1-p([0,f])$.   This shows the probability of a positive return is a function of conditional volatility $\sigma_{t+1|t}$ and so as $\sigma_{t+1|t}$ increases, the probability of a positive return falls.    In order to be able to benefit from this relationship a forecast for $\sigma_{t+1|t}$ is required.

Much literature exists on the subject of volatility forecasting, a summary of which is beyond the scope of this paper so instead the reader is directed to three excellent reviews \citep{Granger2003a, pesaran2009model, zaffaroni2008large}.  The main finding of these reviews is that the sophisticated volatility models can not out perform the simplest models with any statistical significance and for that reason we use the IGARCH(1,1), otherwise known as the J.P. Morgan Risk Metrics EWMA model \citep{JPM1996a, Kondor2001a}.   A drawback to this approach are the recent findings in the literature that volatility estimation with data above $\sim$20-minute frequency can lead to artifacts in the estimate \citep{Andersen2011a}.


The EWMA methodology exponentially weights the observations, representing the finite memory of the market, as per Equation \eqref{ewmaEq},
\begin{eqnarray}
    \sigma_{t+1 | t}  & = & \sqrt{(1 - \lambda) \sum_{\tau = 0}^{\psi}    \lambda^{\tau} \Delta y_{t - \tau}^{2}}
    \label{ewmaEq}
\end{eqnarray}
The model has two parameters $\psi$ (window size) and $\lambda$ (variance decay factor where $0 < \lambda < 1$) which are fixed a-priori with a trade-off between $\lambda$ and $\psi$, with a small $\lambda$ yielding similar results to a small $\psi$. The original J.P. Morgan documentation suggests using $\lambda = 0.94$ with daily frequency data, though we increase the reactivity of the term to fit our one minute frequency data and set $\lambda = 0.79$ \citep{Pesaran2007a, Patton2010a}.  This leaves the only parameter of the model as the number of historical observations $\psi$ to include in the estimate.

There are many technical indicators that are based on volatility in the popular trading literature, including bollinger bands, the ratio of implied to realized volatility, and the ratio of current volatility to historical volatility. We choose to implement the latter termed the \emph{volatility ratio} as designed by Chande in 1992 \citep{Chande1992a} which requires estimating conditional volatilities for ``now'' and in the ``past'' \citep{Colby2002a, volRatInvest, volRatInvest2}.    We parameter sweep the ratio and select values $\psi_{fast} = 50$ and $\psi_{slow} = 100$ based on stability and predictive performance.  The input to Algorithm \ref{splineTrading} is given by ${\bf X} = \nicefrac{\sigma_{t+1 | t}(\psi_{\text{fast}})}{\sigma_{t+1 | t}(\psi_{\text{slow}})}$.

\subsection{Predictor II: Seasonality}
\label{SeasonalitySection} 
Seasonality is an extremely well documented effect in the financial markets and is defined by returns from securities' varying predictably according to a cyclical pattern across a time-scale \citep{Bernstein1998a}.  The time-scale of the variation in question varies from multi-year \citep{booth2003a} to yearly \citep{lakonishok1988seasonal}, monthly \citep{ariel1987monthly}, weekly \citep{franses2000modelling}, daily \citep{peiro1994daily} and intraday \citep{taylor2010a}.  The fact that the periodicity (i.e. frequency) is fixed and known a-priori, distinguishes the effect from other cyclical patterns in security returns \citep{Taylor2007a}.

Intra-daily seasonality is where returns vary conditionally on the location within the trading day. Hirsch observes in 1987, that in the case of the Dow Jones Industrial Average, the market spends most of the trading day going down and a very small amount of time going up (i.e. the rises are large and fast and the falls are gradual and slow), with the rises happening post-open and post-lunch \citep{Hirsch1987a}.

A wide range of methodologies for extracting seasonality signals from financial data exist in the literature, including FFT \citep{murphy1987a}, seasonal GARCH \citep{baillie1991a}, flexible Fourier form \citep{Andersen1997a}, wavelets \citep{gencay2002a}, Bayesian auto-regression \citep{canova1993a}, linear regression \citep{Lovell1963a} and splines \citep{Huptas2009a}.  As we know the size of the cycle a-priori, believe the effect to be non-linear and prefer to work in the time-domain, splines are chosen to estimate the relationship between time of day and security return.  The use of splines seems to be a well accepted way of capturing seasonality, for example \citep{martin2005a, Robb1980a, caceres2007a,taylor2010a, martin2010a}.

Following the approach of Martin et al a \emph{seasonal index} is used to quantify the cycle \citep{martin2010a}.  Here the author constructs an index by defining the period of time under consideration and then partitioning it into a periodic grid between one and $T$ and then assigning observations to buckets on this grid. The authors then capture the seasonal variation by fitting a spline to the seasonal index and bucketed-data.   In the case of our one-minute frequency data the size of the period is $T=856$. The input to Algorithm \ref{splineTrading} is given by ${\bf X} = \left[1,\dots, T \right]$.

\subsection{Simulation and Results}
\label{splineSimSection}
\begin{figure*}[!ht]
    \begin{center}
        {\includegraphics[width=5in,keepaspectratio]{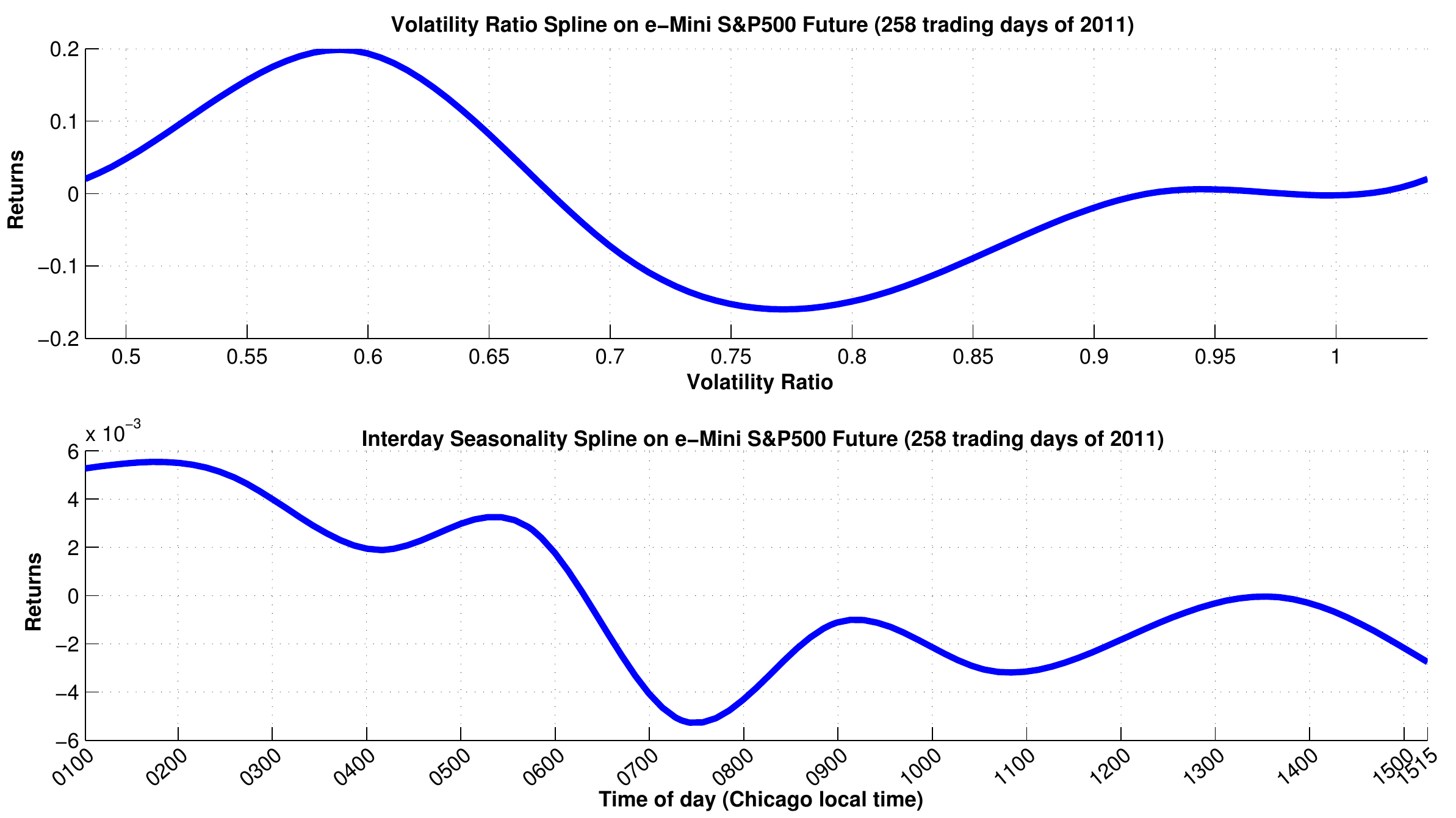}}
        \caption{Forecasting splines. Subplot one shows the spline generated by the volatility ratio predictor. Subplot two shows the spline generated by the seasonality predictor. This approach could be generalized when using ${\cal N}$ predictors, by generating an ${\cal N}$-dimensional spline.}
        \label{skynetSplines}
    \end{center}
\end{figure*}

For our data set consisting of one-years worth of ES data at one minute frequency, Algorithm \ref{splineTrading} is applied to the two predictors and results presented.  Firstly, the two splines generated from the training data set are shown in Figure \ref{skynetSplines}.  It can be seen the relationship is non-linear. It is also clear that the integral of the splines is zero, meaning that a series of random evaluations of the spline will lead to a zero mean signal as required.  By the degree of local structure of the splines it is clear that these are empirical relationships, however this does not invalidate them as predictors, but merely requires a stronger belief in the underlying economic hypotheses behind them.  The economic interpretation of Figure \ref{skynetSplines} for the volatility ratio predictor is that a small (0.6) ratio of recent to old volatility means that risk is falling, and so the spline suggests buying. A large (0.8) ratio of recent to old volatility means that risk is rising, and so the spline suggests selling.  For the seasonality predictor the spline suggests buying in the early morning and selling in the afternoon.\newline
The choice of the number of knots for the spline is important. Too many knots means the spline will be very tightly fitted to the data, while too few knots may fail to capture the relationship of interest. The problem with over-fitting the relationship being that the in-sample performance will be great, but the out-of-sample performance will be poor. Hence it is a matter of balance which is decided upon by intuition about the variability of the underlying economic relationship.   6 knots are chosen for the volatility predictor and 10 knots for the seasonality predictor. Increasing the number of knots on the volatility predictor to 40, doubles the predictive performance in-sample, but is probably just fitting to noise.

The performance of the two strategies can now be simulated against a benchmark of a long-only strategy.  Special care is taken to ensure that the simulation is a truly out-of-sample simulation. Specifically, each one of the data points used to evaluate a trade had not been used in any of the previous stages of model identification, learning and estimation.  The annualized Sharpe ratio is a popular measure of risk-adjusted return and is defined as $\frac{\sqrt{N}(\mu - r)}{\sigma}$ where $\mu$ is the mean strategy return, $\sigma$ is the standard deviation of the strategy return, $r$ is the risk free rate and $N$ is the number of trading periods in the year. The ratio is computed by calculating a vector of daily returns, generated by finding the total intraday strategy return each day and setting $N=258$.  This aggregation approach is preferable to scaling by $\sqrt{N}$ for intraday $N$, as the output is more stable.  As our final signal is zero mean and interest is earned at rate $r$ on short futures positions, we set $r=0$.

\begin{figure*}[!ht]
    \begin{center}
        {\includegraphics[width=5in,keepaspectratio]{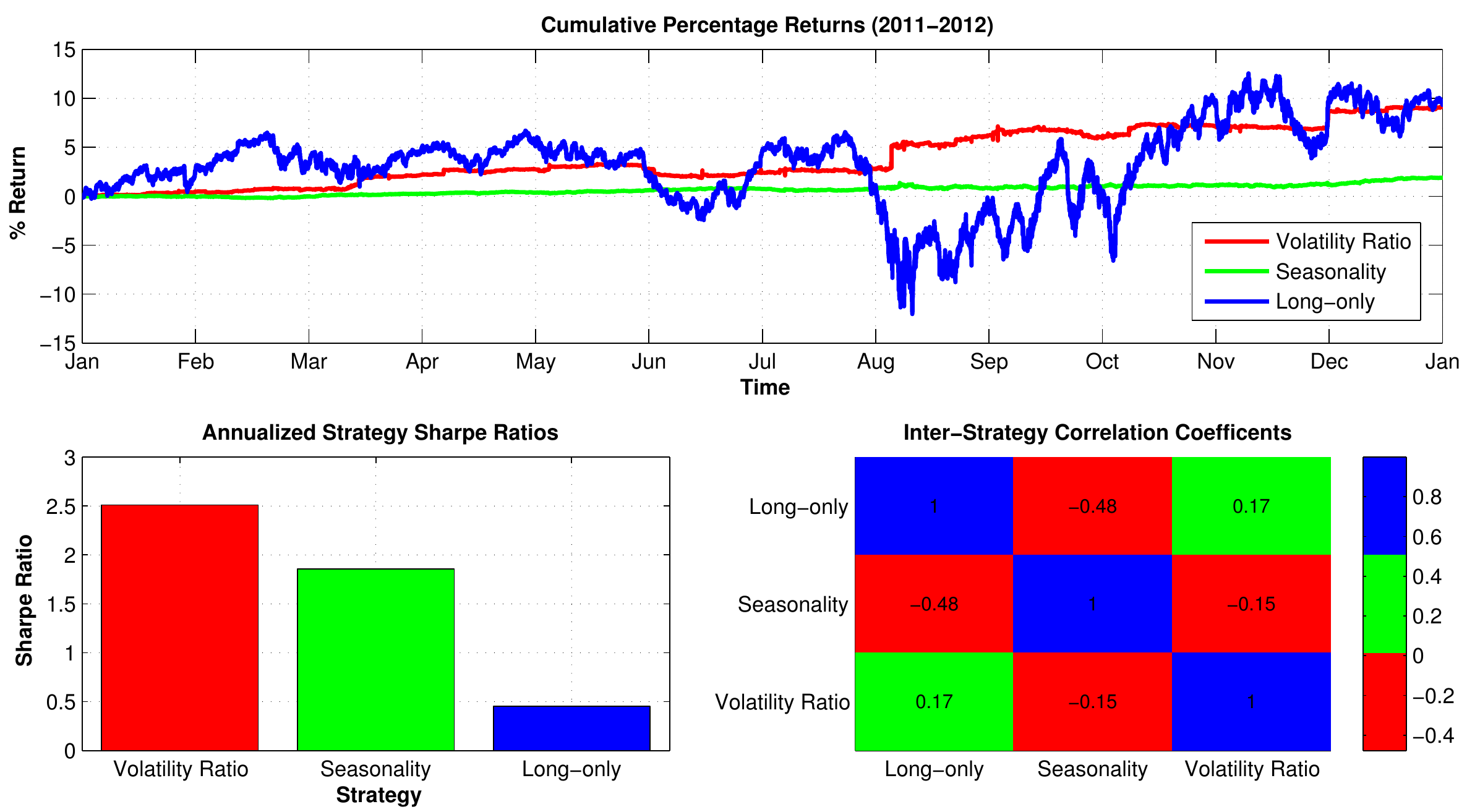}}
        \caption{Forecasting splines results. Subplot one shows the annual returns for the two strategies against a long-only portfolio for the 258 trading days of 2011.  Subplot two shows the mean (pre-cost) annualized Sharpe ratios for the strategies. Subplot three shows the correlation coefficients between the strategies.}
        \label{splinesSimulation}
    \end{center}
\end{figure*}

The results of the simulation are shown in Figure \ref{splinesSimulation}.  Subplot one shows the annual returns for the two strategies against a long-only portfolio for the 258 trading days of 2011.  It can be seen that the returns profile is different for the two strategies so that while the volatility strategy return is higher it also more volatile which results in the two strategies having similar risk adjusted return profiles. Subplot two shows the mean (pre-cost) annualized Sharpe ratios for the strategies.  The Sharpe ratio of both strategies is around 2.0, as commonly required for an intraday trading strategy to be successful.  Subplot three shows the correlation coefficients between the strategies, which are either small and positive or negative, as required for a diverse portfolio.\newline
In summary both predictors seem to have traction against forecasting the returns of ES and thus contain information of predictive use.  For that reason we try and incorporate them into our HMM momentum model.  The ``classical way'' of doing this would be to combine the final three signals, for example, by taking a weighted mean.  Rather than combine the signals outright, the information held in the splines is used in the learning phase.

\section{Learning With Side Information}
\label{learningWithSideInfo}
\subsection{Introduction}
The HMM of Figure \ref{markovM2} states that the probability of transitioning between momentum states is only dependent on the last momentum state, $p(m_t | m_{t-1})$. From Section \ref{sideInfoSection} we have two splines that we know contain useful information when it comes to predicting security returns.     In this Section the HMM is re-specified by incorporating the side information held in the splines, such that the transition distribution is given by $p(m_t | m_{t-1}, x_t)$.    The belief behind this new model is that the extrinsic data is of value to predicting the change in price of the security. Essentially we are saying that not all of the securities' variance can be explained by the momentum effect, even though we believe it to be the dominant factor.

\subsection{Input Output Hidden Markov Models}
\begin{figure}[!ht]
    \begin{center}
        {\includegraphics[width=3.5in,keepaspectratio]{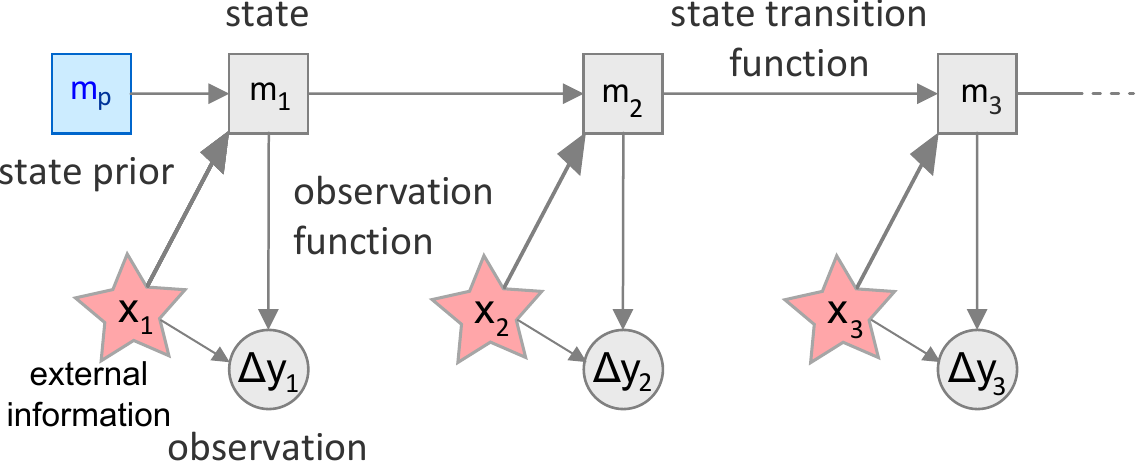}}
        \caption{Bayesian network showing the conditional independence assumption of a synchronous IOHMM. $\Delta {\bf Y}$ is an observable discrete \emph{output}, ${\bf X}$ is an observable discrete \emph{input} and ${\bf M}$ is an unobservable discrete variable.     The model at time $t$ is described by the latent state conditional on the observed state and some external information $p(m_t | \Delta y_{1:t}, x_t)$. }
        \label{newModelMarkov}
    \end{center}
\end{figure}

In  \emph{Input Output Hidden Markov Models} (IOHMMs) the observed distributions are referred to as \emph{inputs} and the emission distributions as \emph{outputs} \citep{bengio1995input}.    Like regular HMMs, IOHMMs have a fixed number of hidden states, however the output and transition distributions are not only conditioned on the current state, but are also conditioned on an observed discrete input value ${\bf X}$.   In the HMM of Section \ref{formingTheModel}, ${\bf \Theta}$ was chosen to maximize the likelihood of the observations given the correct classification sequence $p(\Delta {\bf Y}  | {\bf M}, {\bf \Theta})$. IOHMMs are trained to maximize the likelihood of the conditional distribution $p(\Delta {\bf Y} | {\bf X}, {\bf \Theta})$, where the latent variable ${\bf M}$ is stochastically relayed to the output variable $\Delta {\bf Y}$.  A schematic of the model is shown in Figure \ref{newModelMarkov}.\newline
We consider the simplifying case where the input and output sequences are synchronous \citep{bengio1999markovian}.  Such a system can be represented with discrete state space distributions for emission $p(\Delta y_t | m_t, x_t)$ and transition $p(m_t | m_{t-1}, x_t)$.  When the extrinsic predictor and the HMM momentum predictor have different time stamps, or are of different sampling frequencies, an asynchronous setup is required, adding computational complexity to the forward-backward recursion \citep{bengio1996algorithm}.   It is noted such a technique could allow signals of a lower frequency to be used in a high-frequency inference problem, for example, low-frequency macro-economic data could be used to bias intraday trading.

The literature suggests three main approaches to learning in IOHMM: Artificial neural networks \citep{bengio1995input}, partially observable Markov decision processes \citep{bauerle2011markov} and EM \citep{bengio1999markovian}.  As Baum-Welch (an EM variant) was used for learning in the HMM case, in order to be consistent we opt to learn by EM for the IOHMM case too.   In terms of Algorithm \ref{skynetBW} the only changes required to deal with the IOHMM case are to lines 8 and 13,
\begin{eqnarray}
  \alpha_t(k) &=& \sum_{m_{t-1}}p(z_t|m_t, x_t) p(m_t|m_{t-1}, x_{t-1})\alpha_{t-1}(k) \nonumber\\
  \beta_t(k)&=&\sum_{m_{t+1}} p(z_{t+1} | m_{t+1}, x_{t+1}) p(m_{t+1} | m_t, x_t) \beta_{t+1}(k)\nonumber
\end{eqnarray}
To implement this methodology a different ${\bf A}$ is trained for every unique value of ${\bf X}$.  Such an approach has the drawbacks of over parameterization and requiring large amounts of data. This is solved by discretizing the spline according to its roots, with $R-1$ roots giving $R$ ``buckets'' of spline. $x_t$ is then aligned with $\Delta y_t$, and $\Delta y_t$ assigned to one of the $R$ buckets, the contents of each bucket being concatenated to give a data vector.  Baum-Welch learning with Algorithm \ref{skynetBW} is then carried out on each of these vectors, as before.     As the transition distribution $p(m_t | m_{t-1}, x_t)$ is time sequential, concatenating the bucketed data is strictly incorrect as occasionally $p(m_t | m_{t-\tau}, x_t)$ occurs, where $\tau>1$.   In the case of the two splines in Figure \ref{skynetSplines}, the discretization gives $R=5$ and $R=2$ for the volatility and seasonality predictors respectively. Given the smoothness of the splines, concatenation is rare and so the resulting small loss of Markovian structure can be ignored.  The obvious advantage of discretizing by roots is that parameters $\left\{ {\bf A}_1, {\bf A}_2, \dots, {\bf A}_{R} \right\}$ map to \emph{signed} returns. The learning algorithm for IOHMM is shown in Algorithm \ref{skynetBW3}.
\begin{algorithm}
    \caption{IOHMM Learning. \hspace{0.2cm} \\ ${\bf \hat{\Theta}} = \text{iohhmLearning}(\Delta {\bf Y}, {\bf X})$}
    \begin{algorithmic}[1]
    \label{skynetBW3}
        \STATE $R = \text{NewtonRaphson}\left( {\cal G}\right)$                    \COMMENT{Find the roots of spline ${\cal G}$}
        \STATE ${\bf Z}_{1:R} = \text{map}\left(\Delta {\bf Y}, {\bf X}, R \right)$  \COMMENT{Map $\Delta {\bf Y}$ to buckets corresponding to the roots of  ${\cal G}$}
        \FOR{$r=1$ to $R$}
            \STATE ${\bf \hat{\Theta}}_r = \text{BW}({\bf Z}_r)$  \COMMENT{Baum-Welch on the $R$ buckets, as per Algorithm \ref{skynetBW}}
        \ENDFOR
    \end{algorithmic}
\end{algorithm}

Using the methodology described above, two independent predictions are generated for each of the two IOHMM models, one for the volatility ratio and one for seasonality.  However, it maybe the case we wish to combine the two predictors into a single prediction.  In this case of more than one predictor, ${\bf X}$ is treated as multivariate and a multi-dimensional spline is generated.  Subject to some appropriate discretization of the spline, Algorithm \ref{skynetBW3} can then be applied to solve $p(m_t | m_{t-1}, \b{x}_t)$ where $\b{x}_t$ is a vector.

\section{Inference Phase}
\label{inferanceSection}
We first present inference for the default HMM case and then consider the IOHMM case.  The aim of the inference phase is to find the marginal predictive distribution $p(\Delta y_{t}|\Delta y_{1:t-1}, {\bf \Theta})$. This is found using the forward algorithm \citep{Bishop2006}.

The likelihood vector, size $K \times 1$, corresponds to the observation probabilities and together with the transition probabilities fully describes the model. It is defined as,
\begin{eqnarray}
 p(\Delta y_t | m_t=k, {\bf \Theta})&\propto& \text{Norm}\left(\Delta y_t ; \mu_k,\sigma_k^2 \right) \label{gauss} \\
 &=& \frac{1}{\sigma_k \sqrt{2\pi} } \exp{\left(-\frac{1}{2 \sigma_k^2} \left[\Delta y_t  - \mu_k \right]^2 \right)}\hfill \nonumber
\end{eqnarray}
If the Gaussian assumption of Equation \eqref{PLR} was dropped then Equation \eqref{gauss} would be of a different form. Or in the case of a non-parametric approach, the density of $p(\Delta {\bf Y} | {\bf M}, {\bf \Theta})$ would be evaluated at this step.

The first step of the prediction is different to the subsequent steps, due to not yet being in the recursive chain. The first step starts with a prior over the hidden states,
\begin{eqnarray}
  p(m_1 = k | \Delta y_1) &\propto&  p(m_1=k)  p(\Delta y_1 | m_1=k) \nonumber\\
   &\propto& \pi_k   \times \text{Norm}\left(\Delta y_1 ; \mu_k,\sigma_k^2 \right) \nonumber\\
   &=& \frac{\pi_k \times  \text{Norm}\left(\Delta y_1 ; \mu_k,\sigma_k^2 \right)}{\sum_{k^{'}} \pi_{k^{'}} \times  \text{Norm}\left(\Delta y_1 ; \mu_{k^{'}},\sigma_{k^{'}}^2 \right)}\nonumber
\end{eqnarray}
Once initialization has been dealt with, the rest of the process can be decomposed into a recursive formulation.  The recursions update the posterior \emph{filtering distribution} in two steps: Firstly a prediction step propagates the posterior distribution at the previous time step through the target dynamics to form the one step ahead prediction distribution. Secondly an update step incorporates the new data through Bayes' rule to form the new filtering distribution.     The filtering distribution $\omega_{t | t, k}$ is given by,
\begin{eqnarray}
\omega_{t | t, k} &\triangleq&   p(m_t = k | \Delta y_{1:t})\nonumber \\
             &\propto& p(m_t = k | \Delta y_{1:t-1}) p(\Delta y_t | m_t = k) \nonumber\\
             &\propto& \omega_{t|t-1,k} p(\Delta y_t | m_t = k)\nonumber\\
             &=&\frac{\omega_{t | t-1,k} \times \text{Norm}\left(\Delta y_t ; \mu_k,\sigma_k^2 \right)}{\sum_{k^{'}} \omega_{t | t-1,k^{'}} \times \text{Norm}\left(\Delta y_t ; \mu_{k^{'}},\sigma_{k^{'}}^2\right)}\nonumber
\end{eqnarray}
The \emph{predictive distribution} $\omega_{t | t-1,k}$ is found by multiplying the filtering distribution by the state transition matrix,
\begin{eqnarray}
  \omega_{t | t-1, k} &=& \sum_{k^{'}} a_{kk^{'}} p(m_{t-1}=k^{'} | \Delta y_{1:t-1}) \nonumber\\
                    &=& \sum_{k^{'}} a_{kk^{'}} \omega_{t-1 | t-1,k^{'}}\nonumber
\end{eqnarray}

The prediction $\Delta \hat{y}_t$ is then found by taking the expectation of the marginal predictive density distribution $p(\Delta y_{t}|\Delta y_{1:t-1})$,
\begin{eqnarray}
                \Delta \hat{y}_t   &=& \sum_{\Delta y_t} \Delta y_t \times p(\Delta y_t | \Delta y_{1:t-1})  \nonumber\\
                                   &=& \sum_{\Delta y_t} \Delta y_t \sum_{k} p(m_t = k | \Delta y_{1:t-1}) p(\Delta y_t | m_t = k)  \nonumber\\
                                   &=& \sum_{k} \omega_{t | t-1,k} \times \mu^{*}_k \nonumber
\end{eqnarray}
Where $\mu^{*}_k$ is the mean of the discretized Gaussian $p(\Delta y_t| m_t = k)$. The full approach is summarized in Algorithm \ref{skynetAlg}.
\begin{algorithm}
    \caption{HMM Prediction. \hspace{0.2cm} \\ $\text{Signal} = \text{HMM}(\Delta {\bf Y}, {\bf \Theta})$}
    \begin{algorithmic}[1]
    \label{skynetAlg}
        \STATE {\bf Update for first step}
        \STATE $\omega_{1|1,k} = \frac{\pi_k \times \text{Norm}\left(\Delta y_t ; \mu_k,\sigma_k^2 \right)}{\sum_{k^{'}} \pi_{k^{'}} \times \text{Norm}\left(\Delta y_t ; \mu_{k^{'}},\sigma_{k^{'}}^2 \right)}$
            \FOR{$t=2$ to $T$}
                \STATE {\bf Predict}
                \STATE $\omega_{t | t-1, k} = \sum_{k^{'}} a_{kk^{'}} \omega_{t-1 | t-1, k^{'}}$
                \STATE $\Delta \hat{y}_t = \sum_{k} \omega_{t | t-1,k} \times \mu^{*}_k$
                \STATE
                \STATE {\bf Update}
                \STATE $\omega_{t | t, k} = \frac{\omega_{t | t-1, k} \times \text{Norm}\left(\Delta y_t ; \mu_k,\sigma_k^2 \right)}{\sum_{k^{'}} \omega_{t | t-1,k^{'}} \times \text{Norm}\left(\Delta y_t ; \mu_{k^{'}},\sigma_{k^{'}}^2 \right)}$
                \STATE
                \STATE {\bf Output}
                \STATE $\text{Signal}_{t} = \text{TF}(\Delta \hat{y}_t)$ {\COMMENT{Apply a transfer function}}
            \ENDFOR
    \end{algorithmic}
\end{algorithm}

Inference in the IOHMM case is very similar to the HMM case, though here ${\bf \Theta}$ is conditional on $x_t$. The IOHMM version of the prediction algorithm is summarized in Algorithm \ref{skynetAlg2}.
\begin{algorithm}
    \caption{IOHMM Prediction. \hspace{0.2cm} \\ $\text{Signal} = \text{IOHMM}(\Delta {\bf Y}, {\bf X}, {\bf \bar{\Theta}})$}
    \begin{algorithmic}[1]
    \label{skynetAlg2}
        \STATE {\bf Update for first step}
        \STATE $\omega_{1|1,k} = \frac{\pi_k \times \text{Norm}\left(\Delta y_t ; \mu_k,\sigma_k^2 \right)}{\sum_{k^{'}} \pi_{k^{'}} \times \text{Norm}\left(\Delta y_t ; \mu_{k^{'}},\sigma_{k^{'}}^2 \right)}$
            \FOR{$t=2$ to $T$}
                \STATE ${\bf \Theta} = {\cal F}\left({\bf \bar{\Theta}}, x_t \right)$ {\COMMENT{Parameter lookup table}}
                \STATE {\bf Predict}
                \STATE $\omega_{t | t-1, k} = \sum_{k^{'}} a_{kk^{'}} \omega_{t-1 | t-1, k^{'}}$
                \STATE $\Delta \hat{y}_t = \sum_{k} \omega_{t | t-1,k} \times \mu^{*}_k$
                \STATE
                \STATE {\bf Update}
                \STATE $\omega_{t | t, k} = \frac{\omega_{t | t-1, k} \times \text{Norm}\left(\Delta y_t ; \mu_k,\sigma_k^2 \right)}{\sum_{k^{'}} \omega_{t | t-1,k^{'}} \times \text{Norm}\left(\Delta y_t ; \mu_{k^{'}},\sigma_{k^{'}}^2 \right)}$
                \STATE
                \STATE {\bf Output}
                \STATE $\text{Signal}_{t} = \text{TF}(\Delta \hat{y}_t)$ {\COMMENT{Apply a transfer function}}
            \ENDFOR
    \end{algorithmic}
\end{algorithm}

\subsubsection{Asynchronous Price Data}
In the above form, Algorithm \ref{skynetAlg} supports data which lies on a discrete time grid.   The popularity of such synchronous methodologies in dealing with financial data arises from the computational challenge of dealing with the huge amounts of data generated by the markets.   In reality, financial data is \emph{asynchronous} due to trades clustering together \citep{dufour2000time}.   Aggregation is the process of moving from asynchronous to synchronous data and this acts as a \emph{zero-one} filter. Such a rough down-sampling procedure means potentially useful high-frequency information is thrown away. The Bayesian approach to this problem is to keep as much information as possible and then let the model decide how what parts are/are not needed.

Our model can be altered to deal with asynchronous data by modifying the observation equation in Equation \eqref{PLR} by scaling up the observation inter-arrival times,
\begin{eqnarray}
    \Delta y_{t_i} = \mu_{m_{t_i}} \Delta t_{i} + \epsilon_{t_{i}}, \hspace{1cm} \epsilon_{t_{i}} \sim N(0, \sigma_{m_{t_i}}^2\Delta t_i)\nonumber
    \label{PLR1a}
\end{eqnarray}
where $\Delta t_i = t_i - t_{i-1}$ is the time between asynchronous observations.  Such a representation suffers the drawback that $\mu_{m_{t_i}}$ does not change evenly over time, but changes asynchronously according to observation time. The HMM could be further modified to incorporate smooth $\mu_{m_{t_i}}$ change, for example by using continuous-time HMMs, but this is beyond the scope of this paper.

\section{Data and Simulation}
\label{dataSimSection}
Data from the CME GLOBEX e-mini S\&P500 (ES) future is used, one of the most liquid securities' in the world. Tick data is used for the period 01/01/2011 to 31/12/2011, giving 258 days data. The synchronous form of the algorithm is implemented and the tick data pre-processed by aggregating to periodic spacing on a one minute grid, giving 856 observations per day. Only 0100-1515 Chicago time is considered, Monday-Friday, corresponding to the most liquid trading hours. 1515 Chicago time is when the GLOBEX server closes down for its maintenance break and when the exchange officially defines the end of the trading day. Only use the front month contract is used, with contract rolling carried out 12 days before expiry.  Only GLOBEX (electronic) trades are considered, with pit (human) trades being excluded. No additional cleaning beyond what the data provider has done is carried out.

\begin{figure*}[!ht]
    \begin{center}
        {\includegraphics[width=5in,keepaspectratio]{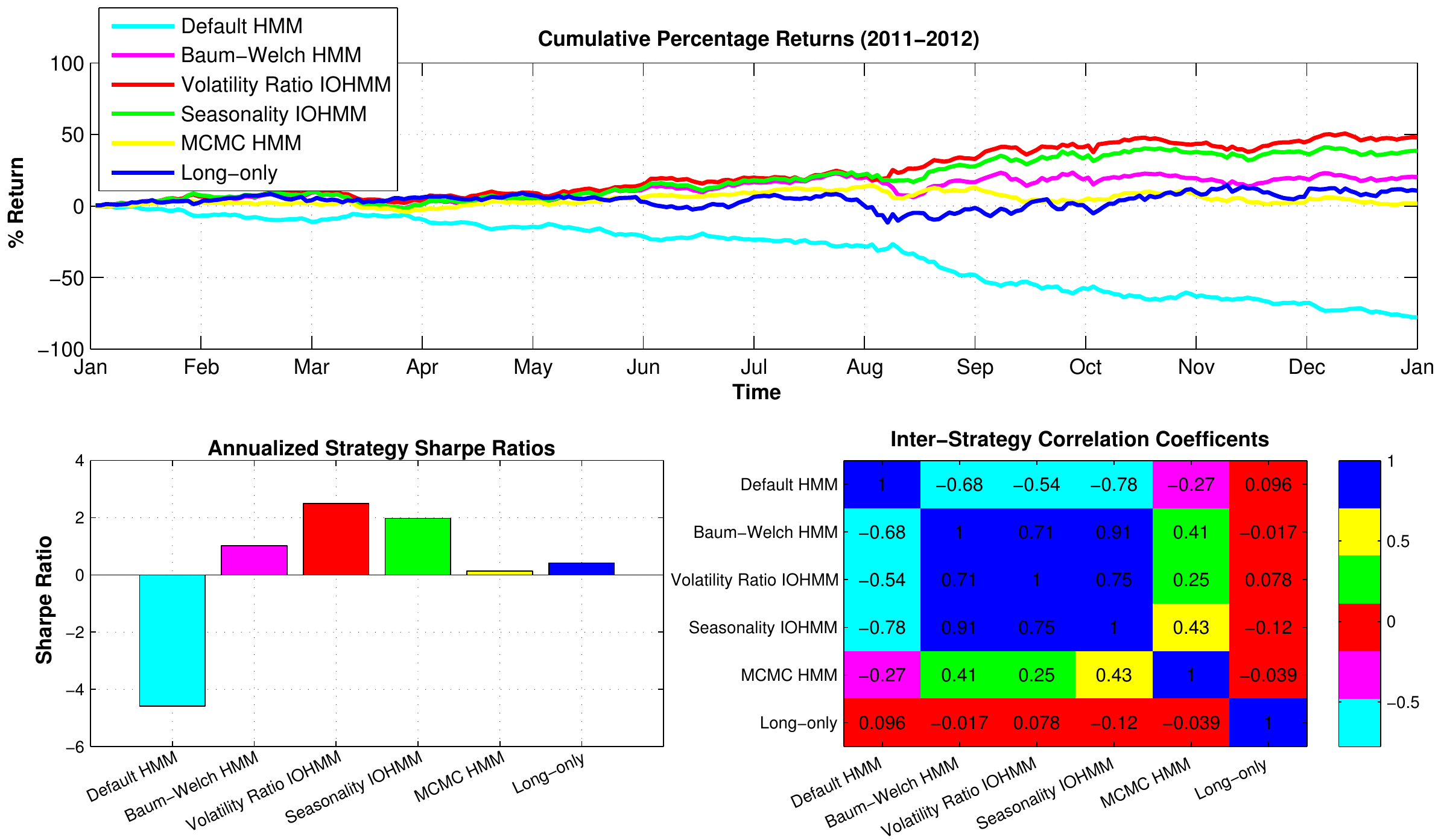}}
        \caption{Simulation results for the five variations of the HMM intraday momentum trading strategy, with $K=2~\text{or}~3$, plus the long-only case.}
        \label{simulationResults00}
    \end{center}
\end{figure*}

As synchronous prices are generated on a close-to-close basis, in simulation the forecast signal is lagged by one period so that look-ahead is not incurred. The \emph{strategy return} is then equal to the security return multiplied by the lagged signal. Learning is carried out on data from the second half of 2010. For all five momentum strategies, the mean and variance were specified for each state (i.e. the system was not tied).  Evaluation of trading strategies is an extensive field, e.g. \citep{Aronson2006a} and so just the key metrics of Sharpe ratio and returns are presented.  The HMM strategies are benchmarked against the long-only case. The pre-cost results of the simulation using ES for the year 2011 are shown in Figure \ref{simulationResults00}.\newline
The performance of the default HMM is the worst of the group of models. This is as expected and reflects the fact that ${\bf A}$ contains no information about the market, as all states are equally likely.  The poor PLR performance can also be explained by the pair of negative trend terms ($\mu_{\text{PLR}} = [-8.99, -0.0207]$), in what was a rising market over the simulation period.  While Baum-Welch was able to beat both the default HMM and the long-only case, MCMC was not able to beat the long-only case. There is no reason why Baum-Welch should be able to outperform MCMC - we believe this may reflect the difficulty in using MCMC correctly.  Reasons for the poor MCMC performance are now discussed, along with suggestions for improvement,
\begin{itemize}[noitemsep,nolistsep]
\item Just as EM can fail to find the true global maxima, MCMC can fail to converge to the stationary distribution of the posterior probabilities \citep{gilks1996markov}.  Common causes for convergence failure are too few draws and poor proposal densities \citep{kalos2008monte}. Ergodic averages of MCMC draws which were generated by random permutation sampling are used to check convergence.  Convergence can be seen to occur in Figure \ref{mcmc_Results} as the entire MCMC chain is roughly stationary for first and second moment parameters. Cowles et al recommend checking for convergence by a combination of strategies including applying diagnostic procedures to a small number of parallel chains, monitoring auto-correlations and cross-correlations \citep{cowles1996markov}. However, we do not believe convergence has failed in this case.
\item The mean emission parameters are Baum-Welch $\mu_{1:3} = [-0.0198, -0.00573, 0.0183]$, MCMC $\mu_{1:3} = [-0.122, -0.0117, 0.121]$. It can be seen that both have negative/zero/positive trend terms, but that the numerical values for the first and third state are quite different. It maybe the case that MCMC has failed to visit all the highly probable regions of the parameter space because of local maxima in the posterior distribution.
\item The step of moving from the distributional estimate to the point estimate presents an opportunity for selecting sub-optimal ${\bf \Theta}$. Our implementation of MCMC approximates the posterior mode by keeping the sample with the highest posterior probability. It is possible however, that this approach could end up selecting a local maxima, as opposed the global maxima, leading to a sub-optimal estimate of ${\bf \Theta}$.  In future work this step could be done by estimating the likelihood of each sample and then taking the maximum.
\item Choice of prior maybe more influential than might be expected \citep{fruhwirth2008comment}. While we have followed the recommendations of the literature, it might be that using a more diffuse prior on ${\bf A}$ would give better results, as it would allow the parameter space to be more thoroughly searched. Failure to search correctly could happen if the existing prior was too strong and overwhelms the data, but this would be unusual given the amount of data used for learning. In particular we believe the use of a uniform prior should cause the results of MCMC and EM to converge. Another approach would be to initialize MCMC with Baum-Welch. If the search moves away from the initial search space, then it might be the case, that the chains are taking a very long time to mix.
\item The proposal density maybe poorly chosen leading to acceptance rates which are too high or too low. In future work we suggest modifying the proposal density to incorporate work from optimal proposal scalings \citep{neal2006optimal} and adaptive algorithms \citep{levine2006optimizing} to attempt to find good proposals automatically.
\end{itemize}
Interestingly the MCMC and Baum-Welch strategy returns have a reasonably high correlation at 0.41, suggesting that they maybe picking up the same market moves, but with MCMC doing so in a less timely (optimal) fashion.  Over the trading times considered, ES rose resulting in a Sharpe ratio of 0.4, approximately equal to its long run average. The failure of MCMC to beat the long-only case, while Baum-Welch does, again points to the fact that parameter selection has failed for MCMC. All of the HMMs have a low correlation to the long-only case, which is as expected given all the HMMs have a zero mean signal. Post-cost results reduce the Sharpe ratio of the HMM strategies by approximately 15\%.

The IOHMM models are both able to beat the Baum-Welch HMM model Sharpe ratio by more than 10\%, reflecting the fact that the model is able to use the information ${\bf X}$, which as known from Section \ref{sideInfoSection} has predictive value.   The Sharpe ratio from the IOHMM models is smaller than that from the individual side-information ${\bf X}$ predictors, because the covariance between the individual predictors and the momentum signal is greater than zero. Even though the Sharpe ratio of the IOHMM signal is less than the Sharpe ratio of the individual ${\bf X}$ predictors, this is not necessarily a bad thing as the correlation of the IOHMM returns has decreased relative to the benchmark returns.     Institutional investors tend to run ``portfolios of strategies'' in order to diversify strategy risk. Here any strategy with a positive expectation and a low correlation to the existing return stream, maybe worthy of inclusion in that portfolio, even if the performance of the new strategy does not beat the benchmark.   The simulation results shown in Figure \ref{simulationResults00} suggest that this strategy could be worthy of inclusion in such a portfolio.

\section{Conclusions and Further Work}
\label{condAndFW}
\subsection{Conclusions}
This paper has presented a viable framework for intraday trading of the momentum effect using both Bayesian sampling and maximum likelihood for parameters, and Bayesian inference for state.  The framework is intended to be a practical proposition to augment momentum trading systems based on low-pass filters, which have been use since the 1970's. A key advantage of our state space formulation is that it does not suffer from the delayed frequency response that digital filters do.  It is this time lag which is the biggest cause of predictive failure in digital filter based momentum systems, due their poor ability to detect reversals in trend at market change points.

As the number of latent momentum states in the market data is never known, it has to be estimated. Three estimation techniques are used, cross-validation, penalized likelihood criteria and MCMC bridge sampling. All three techniques give very similar results, namely that the system consists of 2 or 3 hidden states.

Learning of the system parameters is principally carried out by two methods, namely frequentist Baum-Welch and Bayesian MCMC. Theoretically MCMC probably should be able to outperform Baum-Welch, however when carrying out simulations on out of sample data, it is found that Baum-Welch gives the best predictive performance. The reasons for this are unclear, but it maybe because selecting a good prior is hard for our system, or that the single point estimate of Baum-Welch maybe close to the ``correct'' value, giving superior performance over the Bayesian marginalization of the parameters by MCMC.

Often a trend-following system will want to incorporate external information, in addition to the momentum signal, leading to the signal combination problem. An IOHMM is formulated as possible solution to this problem. In an IOHMM, the transition distribution is conditioned not only on the current state, but also on an observed external signal. Two such external signals are generated, seasonality and volatility ratio, both with positive Sharpe ratios, and are incorporated into the IOHMM. The performance of the IOHMM can be seen to be improved over the HMM, suggesting the IOHMM methodology used is a possible solution to the signal combination problem.

In addition to presenting novel applications of HMMs, this paper provides additional support for the momentum effect being profitable, pre- and post-cost, and adds to the substantial body of evidence on the effect. While much of the existing literature shows that the momentum effect is strongest at the 1-3 month period, we have shown the effect is viable at higher trading frequencies too.

Finally it is noted that this work is an instance of unsupervised learning under a single basic generative model.   As such it can be linked to other work in the field by noting when the state variables presented in this model become continuous and Gaussian, the problem can be solved by a Kalman filter and when continuous and non-Gaussian the problem can be solved by a particle filter, for example \citep{christensen2012forecasting}.

\subsection{Further Work}
In future work we would like to explore in detail why learning ${\bf \Theta}$ by MCMC results in poorer performance than by Baum-Welch. In particular the selection of the prior and the proposal density seem worthy of further investigation, as discussed in Section \ref{dataSimSection}.

In this paper just the best sample of ${\bf \Theta}$ was retained. An improved prediction might be possible by retaining all the samples and averaging their predictions.  Fully Bayesian inference uses the distributional estimate of ${\bf \Theta}$ output from MCMC. Denoting the training data as ${\bf Z}$ and the out of sample data as ${\Delta {\bf Y}}$, MCMC gives $i = 1,\dots,I$ samples from the posterior distribution, s.t. ${\bf \Theta}_i \sim p({\bf \Theta} | {\bf Z}, {\cal M}_k)$. The predictive density can then be determined by,
\begin{eqnarray}
  p({\Delta {\bf Y}} | {\bf Z}, {\cal M}_k) &=& \int p({\Delta {\bf Y}}| {\bf Z}, {\bf \Theta}, {\cal M}_k)  p({\bf \Theta} | {\bf Z}, {\cal M}_k) d{\bf \Theta} \nonumber\\
   &\approx& \sum_{i=1}^{I} p({\Delta {\bf Y}}| {\bf Z}, {\bf \Theta}_i, {\cal M}_k) \nonumber
\end{eqnarray}
A closely related approach that could also be investigated is Bayesian Model averaging (BMA) \citep{hoeting1999bayesian}. While the Bayesian inference just described performs averaging over the distribution of parameters ${\bf \Theta}$, BMA performs averaging at the level of the model ${\cal M}_k$. BMA might be a sensible approach given the similarity of the MCMC marginal likelihoods used for model selection.

Predictive performance may also be improved by removing the model's parametric assumption and changing to use asynchronous data.  By using a more natural description of emission noise, the fit of the model could be improved. In the current downsampling of the data it maybe that useful high-frequency information is getting thrown away. Using asynchronous data would be the most Bayesian approach, allowing the model to decide what to do with that high-frequency information.

Finally, an interesting area of future research could be to compare the IOHMM methodology with other approaches to signal combination, such as a weighted mean of the Baum-Welch HMM and the individual predictor signals.

\section{Acknowledgements}
We acknowledge use of the following MATLAB toolboxes; Kevin Murphy's ``Probabilistic Modeling Toolkit'' \url{https://github.com/probml/pmtk} and $\text{Sylvia Fr$\ddot{\text{u}}$hwirth-Schnatter\text{'}s}$ ``Bayesf'' \url{www.wu.ac.at/statmath/en/faculty_staff/faculty/sfruehwirthschnatter}.


\bibliographystyle{elsarticle-harv}
\renewcommand{\bibname}{References}
\bibliography{../arXiv/SKYNET}
\end{document}